\documentclass{pasa}%

\usepackage{lscape}

\title[Protoplanetary disk's Rosetta Stone]{A ``Rosetta Stone" for protoplanetary disks: \\The synergy of multi-wavelength observations}
\author[Sicilia-Aguilar et al.]{A. Sicilia-Aguilar$^1$, A. Banzatti$^2$, A. Carmona$^3$,  T. Stolker$^4$, M. Kama$^5$, I. Mendigut\'{i}a$^6$, A. Garufi$^7$, K. Flaherty$^8$, N. van der Marel$^{9}$, J. Greaves$^{10}$\\
\affil{$^1$SUPA, School of Physics and Astronomy, University of St Andrews, North Haugh, KY16 9SS, St Andrews, UK}%
\affil{$^2$Space Telescope Science Institute, 3700 San Martin Drive, Baltimore, MD 21218, USA} %
\affil{$^3$Universit\'e de Toulouse, UPS-OMP, IRAP, 14 avenue E. Belin, Toulouse, F-31400, France}%
\affil{$^4$Anton Pannekoek Institute for Astronomy, University of Amsterdam, Science Park 904, 1098 XH Amsterdam, The Netherlands}%
\affil{$^5$Leiden Observatory, Leiden University, PO Box 9513, 2300 RA, Leiden, The Netherlands}%
\affil{$^6$School of Physics and Astronomy, University of Leeds, Woodhouse Lane, Leeds, LS2 9JT, UK}%
\affil{$^7$Universidad Auton\'{o}noma de Madrid, Dpto. F\'{i}sica Te\'{o}rica, M\'{o}dulo 15, Facultad de Ciencias, Campus de Cantoblanco, E-28049 Madrid, Spain}%
\affil{$^8$Van Vleck Observatory, Astronomy Department, Wesleyan University, 96 Foss Hill Drive, Middletown, CT 06459}%
\affil{$^9$Institute for Astronomy, University of Hawaii, Honolulu, 2680 Woodlawn Drive, Honolulu, HI 96822-1839, USA}%
\affil{$^{10}$School of Physics \& Astronomy, Cardiff University, 4 The Parade, Cardiff CF24 3AA, UK }%
}
\jid{PASA}
\doi{10.1017/pas.\the\year.xxx}
\jyear{\the\year}

\usepackage[authoryear]{natbib}
\bibpunct{(}{)}{;}{a}{}{,}
\usepackage{aas_macros}
\usepackage{hyperref} 
\hypersetup{colorlinks,citecolor=blue,linkcolor=blue,urlcolor=blue}
\begin{document}%

\begin{abstract}
The recent progress in instrumentation and telescope development has brought us different ways 
to observe protoplanetary disks, including interferometers, space missions, adaptive optics, 
polarimetry, and time- and spectrally-resolved data. While the new facilities have changed the way we 
can tackle the existing open problems in disk structure 
and evolution, there is a substantial lack of interconnection between different observing techniques 
and their user communities. Here, we explore the complementarity of some of the state-of-the-art 
observing techniques, and how they can be brought together in a collective effort to understand how 
disks evolve and disperse at the time of planet formation. 

This paper was born at the ``Protoplanetary Discussions" meeting in Edinburgh, 2016. 
Its goal is to clarify where multi-wavelength observations of disks converge in 
unveiling disk structure and evolution, and where they diverge and challenge our current understanding. 
We discuss caveats that should be considered when linking results from different observations, or when 
drawing conclusions based on limited datasets (in terms of wavelength or sample). We focus on disk 
properties that are currently being revolutionized by multi-wavelength observations. Specifically: 
the inner disk radius, holes and gaps and their link to large-scale disk structures, the disk mass, and the 
accretion rate. We discuss how the links between them, as well as the apparent contradictions, 
can help us to disentangle the disk physics and to learn about disk evolution.
\end{abstract}
\maketitle%
%


\section{Introduction \label{intro}}

Protoplanetary disks are both a by-product of star formation and the building blocks of planetary systems.
Formed by gas and dust in an initial proportion of
100:1 \citep{bohlin78,savage79}, their structure and evolution is driven by several interrelated physical mechanisms, including
viscous evolution \citep{hartmannetal98}, magnetospheric accretion \citep{koenigl91}, 
photoevaporation \citep{clarke01}, grain growth \citep{beckwith90,miyake93}, 
dust settling \citep{dalessio99}, and eventually, formation of planetary systems.
Although assuming typical disk lifetimes of a few Myr \citep{sicilia06a,hernandez07,WilliamsCieza2011} is widely accepted,
our understanding of the way disks evolve is still highly uncertain. Moreover, recent observations \citep[e.g. HL Tau;][]{ALMA2015} show that
disk structure and evolution are intimately linked and need to be addressed together: signs that were
previously considered as unmistakable evidence of 
evolution (i.e. dust gaps) may be so common and appear so early, that they may be rather considered as typical 
disk structures.

One of the main problems in understanding disks is that the observable footprints of the diverse disk physics
are highly degenerated, especially, when the available observations
span few wavelengths, or are spatially unresolved.  Different observations 
trace different parts of the disk, which is an additional difficulty for their interpretation.
In addition, disks are physically situated somewhere between
stellar atmospheres and molecular clouds, concerning
densities and temperatures. Densities in disks span at least 10 orders of magnitude, and
temperatures range from about 10K to 10000K, so even well-tested 
theories cannot be easily applied. This is why there is no alternative
to the analysis of multi-wavelength, multi-telescope data.

Protoplanetary disks and their evolution forming planets
cannot be captured in their entirety by looking at details seen at a single wavelength.
Multiwavelength data can help 
breaking the degeneracies, but observing time and sensitivity constraints impose strong limitations on the disks
that can be observed.
Observers usually choose one of two directions: either studying one object in great detail,
or studying statistically significant samples of disks at well-selected (usually unresolved)
wavelengths. High-resolution observations of
bright systems unveil the disk structure and are a key to demonstrate the kind of physical processes
that we can expect in disks, but they only have access to a few, nearby, bright objects,
which may not be representative of most disks, nor solar analogs. Statistically significant observations of
large numbers of disks are needed to reveal the common trends and prevalence of different
disk structures, together with the time evolution, although the lesser detail carries the risk of 
always leaving an underlying degeneracy and it also overlooks object-to-object differences.

\begin{figure*}
\begin{center}
\includegraphics[width=15cm]{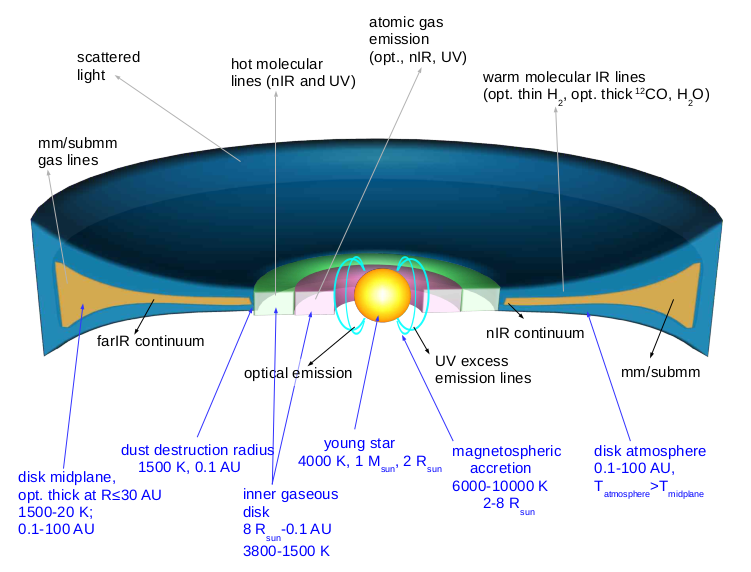}
\caption{A cartoon of the observations and the parts of the disk that they trace, taking as example a young solar analog. 
Although observations trace very different regions and processes in the
disk, we need to keep in mind that they are all connected through the disk itself. 
Note that the complexity of the disk is highly reduced for clarity (for instance, not all the tracers become optically thick at the same location/depth).
For a similar figure regarding the phyiscal processes, see \citet{haworth16}, Figure 1.  Not to scale.}
 \label{cartoon-fig}
\end{center}
\end{figure*}

Current instrumentation (including multi-object capabilities and higher sensitivity on space-
and ground-based facilities) are eroding the separation between individual-system studies and 
statistically-significant observations by improving detectability and time-efficiency, but observational communities 
are still often working apart. This paper aims to determine what can be done and what would be
possible in the near future in terms of observing and understanding protoplanetary disks.
The Disk Rosetta Stone involves observational decryption of disks: sometimes we observe the same phenomenon, but 
use different ``languages" (different wavelengths) to explore the physics. The apparently disconnected
observations are part of the bigger picture  (Figure \ref{cartoon-fig}).

By putting together our observational knowledge, we present a common effort to trace the
structure and evolution of protoplanetary disks using available telescopes and instrumentation,
and discuss how new observing possibilities can be applied in the future to resolve 
the physics and structure of protoplanetary disks around T Tauri stars (TTS) and Herbig AeBe (HAeBe) stars. This paper concentrates on some 
of the most accessible, powerful, and complementary observational techniques currently available.
It is thus not complete regarding all possible observations, it does not include future facilities, 
and it also does not discuss disk chemistry,
which would require another paper by itself. We also limit our study to disks around Class II objects, leaving
aside Class 0/I disks and post-processed, debris disks. Born at the ``Protoplanetary Discussions" in Edinburgh, 2016, this
paper complements, from the observational point of view, the discussions that also gave rise to \citet{haworth16} from the theoretical
side.
Section \ref{innerdisk-sect} discusses the significance and power of the measurements of the inner disk
radius. Section \ref{gaps-sect} deals with holes and gaps in disks, their detectability,
and their implications for disk evolution. The tracers of disk mass 
are explored in Section \ref{diskmass-sect}. Mass accretion is discussed in Section \ref{accretion-sect}.
Variable phenomena, time-dependent processes and disk dynamics 
are presented in Section \ref{timeresolved-sect}. Finally, we include a discussion on the
complementarity and power of the mentioned combined techniques  in Section \ref{discussion-sect}
and our conclusions in Section \ref{conclusions-sect}.

\section{Measuring the inner disk radius \label{innerdisk-sect}}

The first evidence of protoplanetary disks surrounding young stars came from IR excesses, together with
observations of accretion and winds \citep[e.g.][]{strom89}. Given the
wavelengths used in ground-based observations, most of the emission in the near-IR (NIR)
originates in the disk inner rim, being dominated by dust at the dust sublimation radius (T$\sim$1500 K).
The higher densities and shorter orbital periods in the innermost disk led to the prediction 
of inside-out disk dispersal \citep{hayashi85}, later confirmed by the first observations of ``transition disks",
presumed to be in a stage between disked and diskless stars, where the
inner disk rim is larger than the dust sublimation radius \citep{strom89, skrutskie90}.
 
The inner disk, considered as the radial region inwards of $\sim$10-20 AU that produces substantial 
emission in the NIR (in both gas and continuum), is a key region for the formation of
habitable planetary systems, and for Solar Systems analogs.  The large majority of exoplanets discovered to date have semi-major axes within 
$\approx$10\,AU, making this disk region essential for the interpretation of exoplanet data. 

While the inner dusty disk radius is physically set by the sublimation of dust grains at high temperatures (Section \ref{sec: R_in_dust}), 
other processes are expected to take over with time (e.g. grain growth, photo-evaporation, pebble/planetesimal/planet formation), 
pushing it to larger disk radii. The gaseous part of the disk can extend down to the corotation radius or the stellar magnetosphere (a few stellar
radii in size), although depending on the temperature and density, the gas can be molecular or atomic. The hotter inner disk region is a key to understand the onset of disk dispersal through inner gaps in the gas and dust radial distributions.  
We put special emphasis on determining the ``inner disk radius" to constrain the disk
evolutionary stage, noting that the radius  depends on the tracer (gas, dust) used. 

NIR and mid-IR (MIR) observations (1-30$\mu$m) are very sensitive to dust close to the star, due to the large range of temperatures
that produce substantial emission at these wavelengths (ranging from the dust sublimation temperature, $\sim$1500 K, to
$\sim$150 K), and to the large range of dust grain sizes that can produce the excess emission\citep[$\sim$0.1$\mu$m to $\sim$20$\mu$m;][]{miyake93}. 
Hot molecular line observations (e.g. CO, H$_2$) trace the warm molecular layers in the disk.

\begin{figure*}
\begin{center}
\begin{tabular}{c c}
\includegraphics[height=7.3cm]{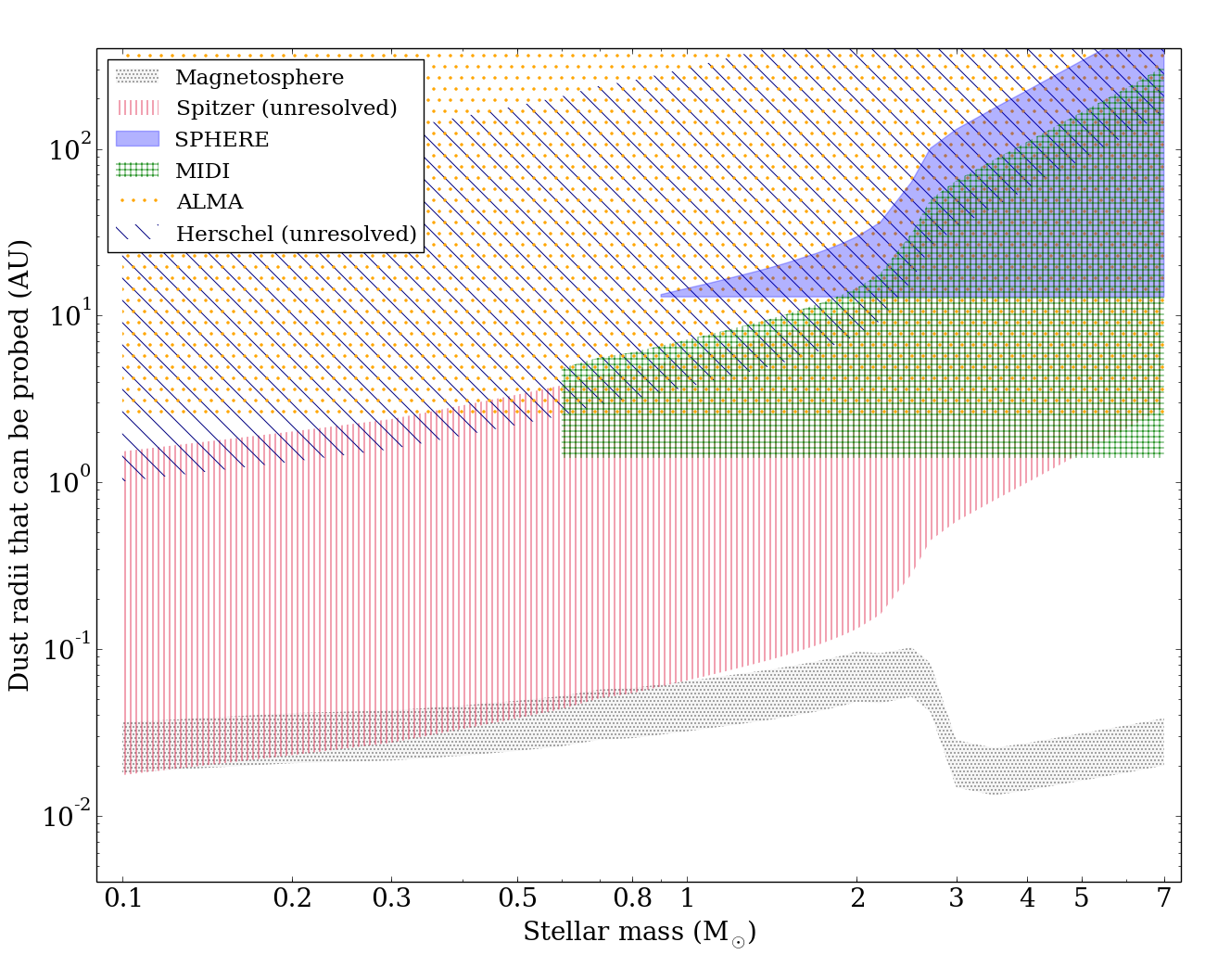} &
\includegraphics[height=7.3cm]{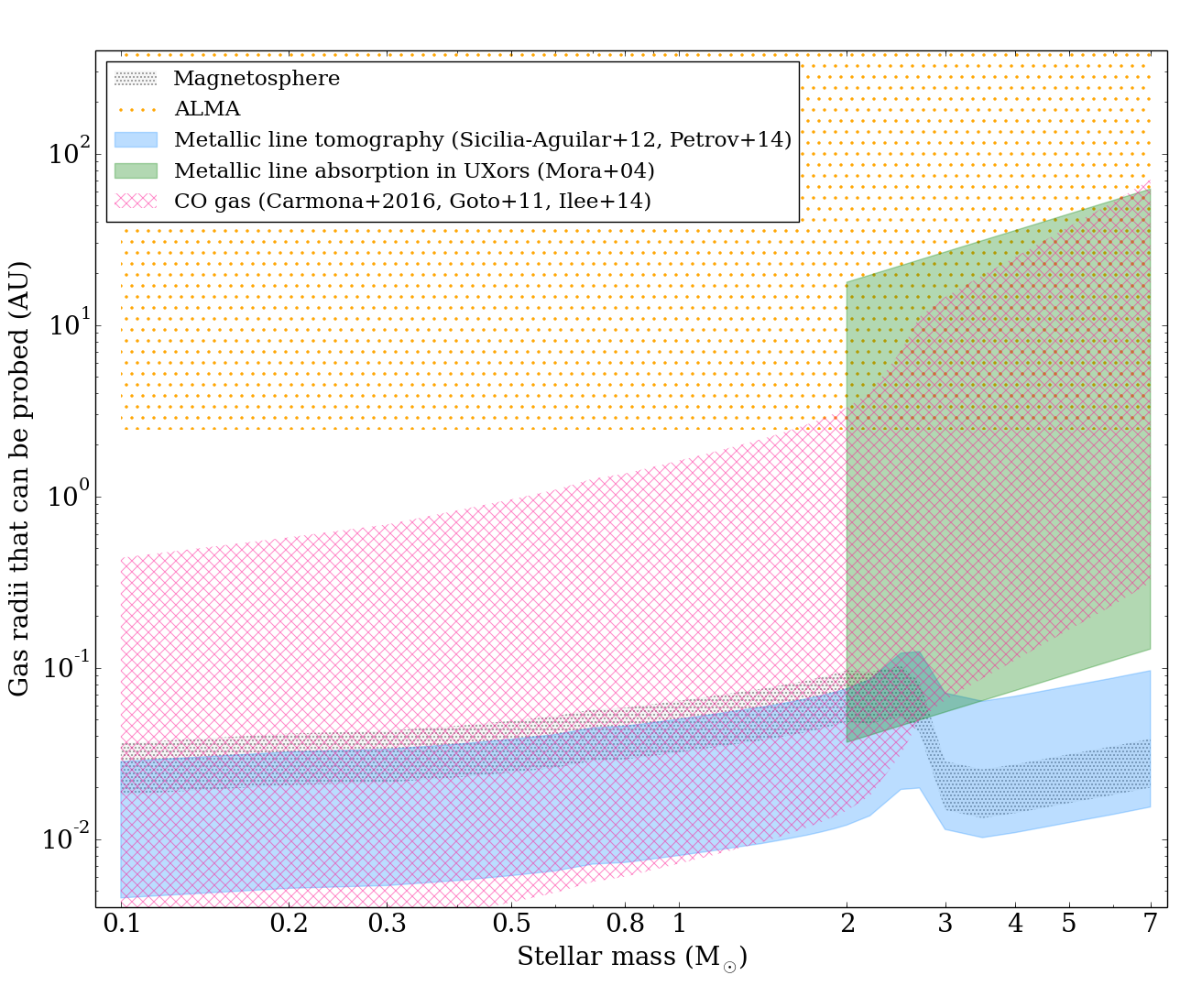} \\
\end{tabular}
\caption{Disk radii that are accessible by different techniques tracing dust (left) and gas (right), for stars with different masses.
The figure shows the regions where different
methods overlap and what they cannot trace. Left: Detectable dust inner disk vs stellar mass, as it can be observed at different wavelengths for an object
at 140 pc distance. Resolved and unresolved
observations are included. For unresolved observations, the detectability depends on the inner rim temperature, and
is subject to model fitting (e.g. SED fitting), so the diagram shows the radii at which dust emission of larger than 15\% over
the photospheric levels can be detected. The lower edge of the observations correspond to the dust destruction radius (T$\sim$1500 K).
For comparison, a stellar magnetosphere (between 4-8 R$_{star}$) is also displayed.
Right: Detectable gaseous inner disk vs stellar mass, as it can be probed
by different techniques. Note that for gas detetion, there is a distinction between atomic gas tracers and molecular gas. Beyond an approximate temperature
of $\sim$2700K, the gas is mostly atomic, although molecular gas can be found up to temperatures $\sim$5000 K, depending on density \citep{Ilee14}.
The CO gas will produce a substantial emission at temperatures $>$300 K (Carmona et al. 2016), although detection may depend on the disk's area. 
Also note that ALMA gas observations at very high resolution are strongly limited by sensitivity, 
so most systems are not expected to be detectable as they do not have enough cold gas so far in.  }
 \label{innerrim-fig}
\end{center}
\end{figure*}

Figure \ref{innerrim-fig} summarizes the parts of the disk that can be detected in gas and dust with various techniques.
The detectability depends on instrumental capabilities: maximum resolution with ALMA\footnote{http://www.eso.org/sci/facilities/alma/documents.html},
limiting magnitudes for SPHERE\footnote{https://www.eso.org/sci/facilities/paranal/instruments/sphere/doc.html} 
and MIDI\footnote{http://www.mpia.de/MIDI/midi\_overview/MIDIoverview.htm}. We also assume that the disk is massive and bright enough,  
on the temperature of the gas or dust in the region. Since the dependency of the temperature
with the radius is very complex (needs to take into account the density, grain sizes, chemistry, structure of the emitting region, scale height, plus
potential heating mechanisms in addition to the central star), we
take a simple approach where the temperature at each radii is assumed to be black-body-like and result from reprocessed star light alone.
Although highly simplified, this figure reveals the main problem when measuring the inner disk rim: different tracers are not sensitive to the
same disk components, and can potentially produce very different results that need to be compared with care.

Here, we consider the inner disk radius (or inner radial extent of the disk) as the radius closer 
to the central star where dust and molecular gas can survive and can be observed 
(the limits of this definition are discussed in each section below). 
In the following Sections, we discuss observations of the inner disk radius of dust and gas
and how the dust sublimation radius (R$_{subl}$) provides the reference to study the onset and evolution of inner disk dust gaps.
We discuss first dust observations of R$_{in, dust}$, then molecular gas observations of R$_{in, CO}$ and the picture emerging from combining the two.

\subsection{Unresolved observations of the inner dusty disk}

NIR observations were the basis of the
first estimates of disk lifetimes \citep{haisch01}.
With the advent of the \emph{Spitzer Space Telescope}, large samples covering most of the
disks and diskless populations in clusters, extended our knowledge of the dusty inner rim over several
AU (see Figure \ref{innerrim-fig}). Spitzer data allowed to conduct statistical studies in disk evolution, including ``transition disks"
\citep[e.g.][]{sicilia06a,najita07,espaillat12} . Despite being unresolved, the number of disks observed during the Spitzer cold mission
is so overwhelming, that the statistical constraints on disk properties (including the presence of inner holes and gaps) and
their lifetimes have provided one of the most complete and general views about the typical structures, dispersal paths, and lifetimes
for the disks around solar- and late-type stars \citep[][among many others]{hartmann05,megeath05,lada06,sicilia06a, hernandez07}.
Such unresolved observations are particularly important to study disk dispersal in solar analogs, since
dispersing disks around low- and solar-mass stars are very often too faint to be resolved otherwise.

Silicate emission is another tracer of the dust grains in the warm disk atmosphere and a signature of the vertical temperature
structure in the disk \citep{calvet92,dalessio06}. Although for most disks the silicate observations are unresolved, it is possible to
obtain spatially resolved silicate data \citep{vanboekel04,juhasz12}. Partly observable from the ground, it was efficiently observed for large
samples of objects thanks to Spitzer/IRS, allowing to study the disk mineralogy in a statistically significant
way. Even though the silicate emission does not
provide information on the global grain properties in the disk, it is a sign of grain processing, heating, mixing, and
transport in the disk. Dust processing happens in all protoplanetary disks, ranging from HAeBe \citep{meeus01,bouwman01,vanboekel05}
to brown dwarf disks \citep{apai05,ricci14}. Crystalline silicates are not found in the ISM \citep{kemper04} and
require very specific conditions for their formation. Thus the mass fraction of crystals and stochiometry of the silicates can be used to trace the
physical conditions in the disk when the material formed, including the
temperature, initial chemical composition, grain sizes, velocities, and the time frame for annealing,
considering the different chemical reactions that give rise to the production of different silicate components 
\citep[e.g. silica, forsterite, enstatite;][]{bouwman01,henning10}. 
The formation of a silicate feature is also strongly
connected to the disk structure, so the size of the emitting region can be estimated even in unresolved
observations, considering the strength of the emission in different silicate bands 
\citep{kesslersilacci07,bouwman08,juhasz10}. 

The silicate feature is optically thin, which allows to estimate the dust mass and composition fraction in the outer disk layers
and, comparing to the continuum, to discern the presence of gaps and holes \citep{bouwman10}
and the presence of large ($\sim$10$\mu$m) grains in the disk atmosphere, used to constrain turbulence and 
settling \citep{sicilia07,pascucci09}. The lack of strong amorphous silicate features in intermediate-aged disks
around M-type stars is interpreted as a sign of grain growth and settling in the innermost disk, which dominates the
10$\mu$m emission \citep{sicilia07}.
The lack of trends between crystallinity and other disk and stellar properties \citep{watson09,sicilia07,sicilia11} 
suggests that their formation depends on many factors, including rapid creation and mixing of crystalline silicates on timescales
much shorter than those of disk evolution \citep{abraham09,juhasz12}. 
Finally, crystalline silicates could be used as indirect signatures of the presence of planets, related to heating by 
shocks \citep{desch05,bouwman10}.

\subsection{Inner extent of dust in disks} \label{sec: R_in_dust}

Although modeling spatially-unresolved spectral energy distributions (SED) provides some constraints 
on the inner disk radius, the most accurate measurements of R$_{in, dust}$ come from spatially-resolved interferometric 
observations of NIR dust emission. These measurements showed that the spatial extent of the hottest dust in disks were 
not consistent with disk models extending up to the star, and were found to correlate with the squared root of the 
stellar luminosity as R$_{in, dust}$ $\propto$ L$_{\star}^{1/2}$ \citep{Monnier02,Dullemond10}. This correlation was 
readily explained with the existence of a dust sublimation front at temperatures of $1300$-$1500\,$K: most of the NIR 
dust emission is emitted by a disk rim located at the dust sublimation front \citep{Natta01,dullemond01}.
The gas inside this rim must be radially optically thin enough to allow a direct irradiation of the rim. 
More physical models of the rim were proposed, including the physics of dust sublimation (Isella \& Natta 2005), 
multiple dust species and full radiative transfer \citep{kama09} and detailed hydrodynamics \citep{flock16}. 
The radial location of the dust rim can provide constraints on the inner disk dust surface density, size distribution, 
and material through an analytical formula or through detailed models \citep{kama09,flock16}.

The advent of new instruments at VLTI, such as AMBER and PIONIER, and the use of very long baselines (CHARA), 
enabled detailed modelling of multi-wavelength 
observations. These showed that additional material (gas or refractory species) could contribute significantly to 
the NIR excess \citep{kraus08,tannirkulam07,benisty10a}. Multi-wavelength observations 
revealed systems with cleared regions \citep{olofsson11,olofsson13,tatulli11,matter14}, 
and optically thin material located inside the gap \citep{kraus12}.  With a larger number 
of observations available, the images of the first AU could be reconstructed, unveiling a complex morphology 
\citep{renard10,benisty10b}.
Although the most detailed studies were focused on a small samples, recent homogeneous PIONIER 
observations of HAeBe stars found that very few objects show well-resolved puffed-up rims, so that the 
sublimation front is rather smooth and not a sharp transition (Lazareff et al. submitted). 
Large-scale emission, contributing to lower visibilities at short baselines, might indicate cavity 
walls. Similar findings were derived from a statistical analysis of MIR observations \citep{millangabet16}.

As a cautionary note, resolved observations are subject to interpretation through models that depend  
on the dust composition and size distribution,
on the structure of the rim, on the dust density, and on the
temperature. The dust composition and size affects the dust opacity, which 
controls the local temperature (see Section \ref{discussion-sect}). Observations of different objects, 
including those with and without inner holes, and combination with multiwavelength observations to track
the gas content and the extended disk structure are a key to put these observations into a broader context.

\subsection{Inner extent of molecular gas in disks} \label{sec: R_in_gas}

Similarly to the hottest dust in the inner disk, the hottest (and innermost, in terms of disk radii) molecular gas 
emission typically peaks in the NIR with ro-vibrational branches at 2--5\,$\mu$m (e.g. Fig.7 in Salyk et al. 2009), 
or in the UV at 1300--1700 \AA\ \citep{france11,france12}.
With due differences, spatial information on the gas-emitting region can be obtained with 
interferometry \citep[][]{eisner10,eisner14}, 
spectro-astrometry \citep[][]{pont08,pont11,brittain15}, position-velocity diagrams 
\citep{goto06,brittain09,vdplas09,carmona11}, and with high-dispersion spectrographs through 
fully spectrally-resolved velocity profiles \citep{brown13,BP15}. 
For optically thin lines, the strength of NIR gas emission lines is linked to the column density, 
temperature, and excitation conditions of hot gas in the inner disk. Optically thick lines
trace the temperature in the emitting region, which has a complex dependency on the gas density. 

The emission from two molecules, H$_2$ and CO, is especially suited to trace the innermost disk region where 
molecular gas can survive: they are abundant (being made by the most abundant atoms), they share a similarly 
high thermal dissociation temperature ($\sim4500$ K) and they can also self-shield to dissociating UV radiation 
to some extent (\citealt{bruderer13}; see Section \ref{sec: in_gaps}). 

The near-IR CO fundamental ro-vibrational 
lines are good tracers of the gas in the inner disk because their energy levels are sufficiently 
populated at the temperatures found at 0.1-10 AU and because their Einstein A coefficients are large, 
making them much stronger than the NIR H$_2$ lines (e.g. Carmona et al. 2008; Bitner et al. 2008), which lack a 
permanent dipole moment. As the CO emission becomes optically thick at low column densities 
(N$_{\rm CO} \sim 10^{16}$ cm$^{-2}$ or N$_H \sim 10^{20}$ cm$^{-2}$), strong lines are formed even 
in the upper regions where the dust is optically thin.
The CO overtone needs higher column densities and temperatures than CO ro-vibrational to be excited 
(N$_{\rm CO} \sim $5$\times 10^{20}$ cm$^{-2}$, T$>$1700 K; Bik \& Thi 2004), being mostly observed in massive young stars.

UV pumping triggers H$_2$ emission at 2.12$\mu$m, which can be detected up to 150 AU if the disk is warm and flared 
(e.g. Carmona et al. 2011).  H$_2$ also has stronger fluorescent transitions (excited by Ly$\alpha$ photons) in 
the UV that been observed with IUE and HST wavelengths and that are also detectable for CTTS and transition disks 
(Valenti et al. 2003; Hoadley et al. 2015). 
The main limitation for these observations
are the brightness of the target and whether there is still enough gas at a high enough temperature, to produce substantial NIR
emission. This poses a general outer limit of $\approx20$ AU to the disk regions that can be studied in the 
thermal NIR, as well as sensitivity limits of the individual techniques. 
All considered to date, high-dispersion spectroscopy of NIR CO emission has provided the largest and most informative 
dataset of molecular gas observations in inner disks \citep{brown13,BP15}.

High-dispersion spectroscopy provides a way to characterize the emitting region of gas in disks even at
scales not directly spatially-resolvable. This is achieved by modeling velocity-resolved 
line profiles broadened by Keplerian rotation in the disk. The observed velocities depend on 
Kepler's law and on the inclination
angle, so that emission from smaller orbital radii has higher velocity shifts. The observed 
line profiles therefore provide measurements of the disk radii where CO is emitting, being
``spatially-resolved" through the velocity shifts. To fully spectrally resolve a line 
from a $\lesssim 10$ AU disk radii, a resolving power of at least 25000
is required, depending on the disk inclination. In the case of NIR observations, this is 
currently achieved by high-resolution slit spectrographs such as CRIRES at the ESO-VLT, iSHELL on the IRTF,
NIRSPEC at Keck, and IRCS at Subaru. This technique has proven to be efficient in surveys 
of velocity-resolved CO emission, especially with the advent of CRIRES 
\citep[e.g.][]{pont11,brown13,BP15}. It allows us to study a radial region between 
0.05-20 AU in 
disks around stellar masses $\gtrsim0.3$ M$_{\odot}$, based on flux sensitivity limits of CRIRES 
of  $\approx2-10^{-16}$ erg cm$^{-2}$ s$^{-1}$
for an unresolved line with resolution 3.3 km/s and a typical line width of
10 km/s.
The minimum column density depends on the assumed gas temperature, the size of the emitting region, and the disk inclination.  
From observations of HD 139614 (Carmona et al. 2016), the minimum column density 
detectable at 3$\sigma$ level was N$_H$ = 5$\times 10^{19}$ cm$^{-2}$ (N$_{CO}$ = 5$\times 10^{15}$ cm$^{-2}$) for a disk of gas between 0.1-1.0 AU at  T=675-1500 K 
inclined 20 degrees around a 1.7 M$_\odot$ star.

If the emitting region is beyond 10--30 AU in disks at 120--140 pc, CO emission lines can be directly spatially 
resolved in position-velocity diagrams \citep{goto06,brittain09,vdplas09,carmona11}. 
If the emitting region is $<10$ AU, information on the spatial scales can be retrieved 
from the spectroastrometry signature of the line profiles in the the 2D spectrum 
\citep[e.g.][]{pont11,brown13,vdplas15}. Spectroastrometry essentially measures the shift on the center of the 
point spread function (PSF) at the position of the emission lines in the 2D spectrum. As the center of the PSF can be measured with an 
accuracy of the order of 1/10 of a pixel with CRIRES (with the assistance of adaptive optics), spatial 
information can be retrieved on spatial scales of the order of 0.01" (= 1 AU at 100 pc), for
bright enough sources. This technique has been successfully implemented in a dozen of disks so 
far (Pontoppidan et al. 2008, 2011; Brittain et al. 2015). Its success depends, among other things, on the
disk inclination and on reaching a good S/N, which strongly limits the observations of faint and low-mass disks.
NIR interferometry has been used to detect CO emission at 2.3\,$\mu$m at disk radii of $<2$ AU \citep{eisner10,eisner14}. The 
low detection rates with this technique suggested that the most effective range to study CO gas in inner 
disks is at 4.6-5\,$\mu$m, where the emission is more frequently found, also for low-mass stars \citep{goto06,brown13,BP15}

All the NIR observing techniques mentioned above provide spatial information on CO emission in the inner disks. 
As they probe the hottest molecular gas, the spatial information is usually regarded as a measurement 
of the smallest stellocentric distance where the molecular gas survives, which we call 
R$_{in, \rm{CO}}$ as being based mostly on observations of CO. These measurements have a
high potential to constrain disk structure and evolution, as we will discuss later.
In velocity-resolved observations, the velocity at the half width at half maximum (HWHM) of the line provides a 
measurement of the disk radius close to where the peak line flux is emitted. At smaller disk radii, less than 
10\% of the line flux is typically emitted, with slight differences depending on how steep line wings are \citep{BP15}. 
In spectroastrometry, R$_{in, \rm{CO}}$ is usually taken at the peak of the spectro-astrometric signal, which 
corresponds to the disk radius that contributes most to the emission \citep[][]{pont08,pont11}. 
Models show that R$_{in, \rm{CO}}$ from CO ro-vibrational emission corresponds to the location of the maximum 
CO intensity when a power law for the intensity, or a power-law column density and temperature profiles are 
assumed. It effectively corresponds to the disk radius where CO emission becomes optically thick. 
From R$_{in, \rm{CO}}$ inward to the star, the column density of CO decreases, so
the value depends on how the temperature increases at lower radii. For a detection, the decline on 
surface density should (combined with the decrease on solid angle) be faster that the increase on temperature. 

As being based on molecular gas, R$_{in, \rm{CO}}$ does not imply that no gas is present closer to the star. 
Atomic gas usually extends inward to the magnetospheric radius to feed stellar accretion, and it is often much easier to detect accretion than
molecular gas, especially in low-mass disks (see Section \ref{accretion-sect}).

\subsection{R$_{in} >$ R$_{subl}$: the onset of inner disk gaps?} \label{sec: in_gaps}
Given that the dust dominates the opacity and controls the amount of UV radiation penetrating the disk, 
the survival of molecular gas is expected to be linked to the presence of shielding dust.  
We would thus expect R$_{in, dust}$ and R$_{in, \rm{CO}}$ to agree with each other in disks, but 
CO is able to self-shield against UV radiation,
surviving down to very low amounts of gas mass even in the total absence of dust grains 
(e.g. $^{12}$CO self-shields at N$_{CO} \sim$10$^{15}$cm$^{-2}$, corresponding to N$_H \sim$10$^{19}$cm$^{-2}$, assuming
standard abundances; van Dishoeck \& Black 1988; Bruderer et al. 2013). Although R$_{in, dust}$ is set by the 
dust sublimation temperature($\sim 1500$ K), R$_{in, \rm{CO}}$ may be smaller than expected 
from the thermal dissociation temperature of CO (4500 K) due to self-shielding. If R$_{in, dust}$ is larger 
than R$_{subl}$, the dust is removed by means other than dust sublimation. In this case, if R$_{in, \rm{CO}}$ 
still matches R$_{in, dust}$ it means that whatever process is removing dust it is also removing CO gas, 
because otherwise CO gas should self-shield and exist down to $<$R$_{subl}$. Therefore, 
comparison of measured R$_{in, dust}$ and R$_{in, \rm{CO}}$ to R$_{subl}$ can help to 
understand the physics and processes that regulate the inner disk structure.

\begin{figure}
\begin{center}
\includegraphics[width=0.5\textwidth]{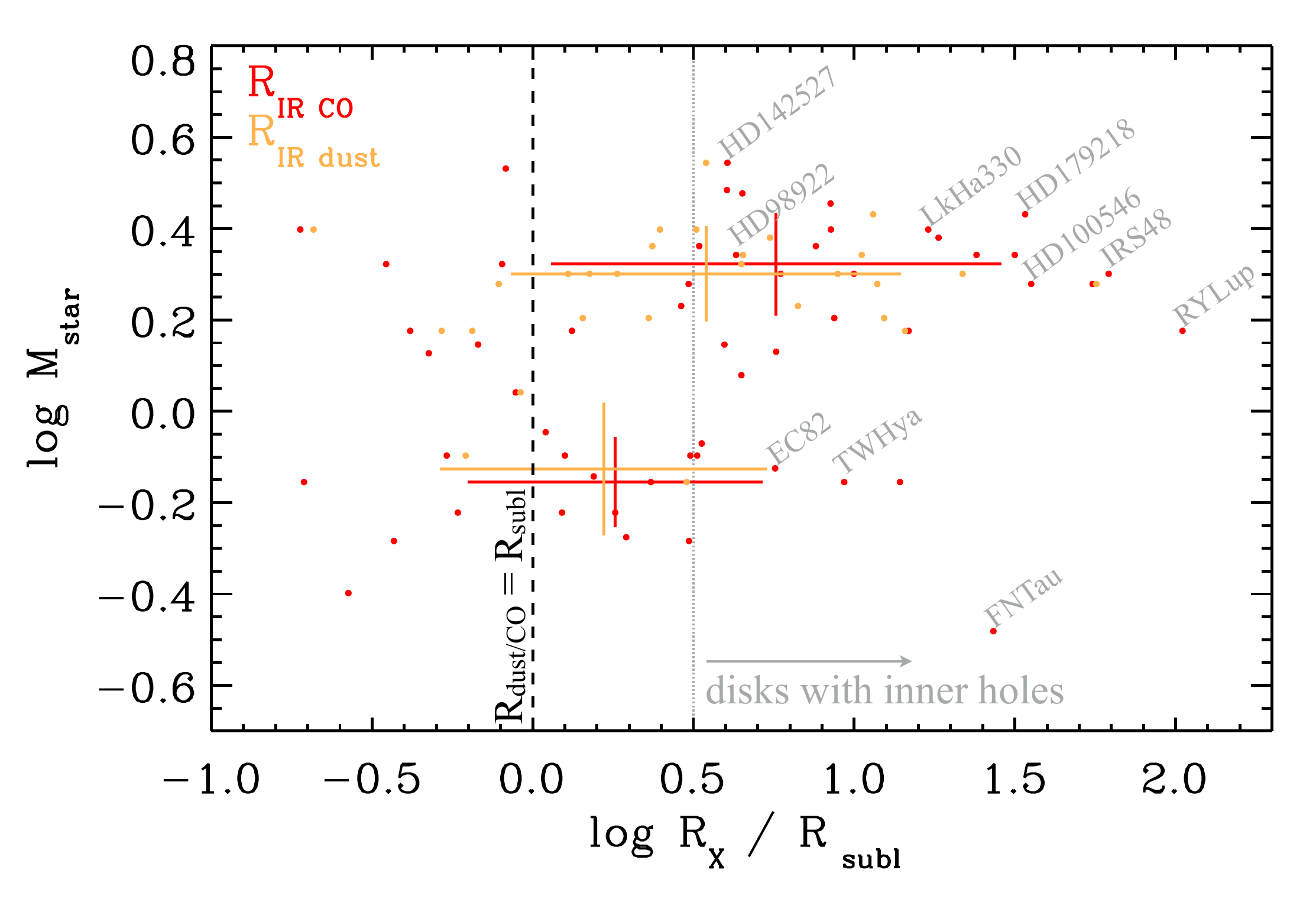}
\caption{Measurements of R$_{IR,dust}$ from IR interferometry \citep[orange points, from][]{anton15,menu15} and of R$_{IR,CO}$ 
from IR spectroscopy of CO gas \citep[red points, from][]{BP15}. The dust/CO radii (R$_X$) are normalized to the dust 
sublimation radii R$_{subl}$ expected from models. Large crosses show median values and median absolute deviations for two stellar mass bins. } \label{fig: Rin_obs}
\end{center}
\end{figure}

In Figure \ref{fig: Rin_obs} we show measurements of R$_{in, dust}$ from IR interferometry (Anthonioz et al. 2015; 
Menu et al. 2015) and of R$_{in, \rm{CO}}$ from IR spectroscopy of CO gas (Banzatti \& Pontoppidan 2015), 
as compared to a simplified parametrization of R$_{subl}$ (Dullemond \& Monnier 2010, Salyk et al. 2011). The two 
probes of R$_{in}$ agree with each other and with R$_{subl}$ in most disks around stars with masses $<1.5$ M$_\odot$,
 supporting the idea that CO survives where dust survives, and that dust survives until it is destroyed by
the high temperatures close to the star. R$_{in, \rm{CO}}$ is much larger than R$_{subl}$ in some disks, though, and 
notably some of them are know to be ``transitional" disks from observations of the inner dust (e.g. TWHya). Remarkably, 
for stellar masses of $>1.5$ M$_\odot$, R$_{subl}$ is smaller than both R$_{in, dust}$ and R$_{in, \rm{CO}}$ in the vast 
majority of disks. This has been recently interpreted as evidence that most of these disks are forming inner gas and 
dust holes by means different than dust sublimation \citep{maask13,BP15,menu15}.  

The combination of observations at different wavelengths can thus help to explore the various possibilities of 
disk dispersal in the inside-out scenario to understand the physics and connections between gas and 
dust, and the process of accretion and transport in the innermost disk, all highly relevant for the 
formation of terrestrial planets and Solar Systems analogs.

\section{Imaging disk gaps and large-scale asymmetries \label{gaps-sect}}

Large-scale asymmetries are increasingly attracting the attention of the disk and planet communities, 
thanks to the recent spatially-resolved images of structures in disks. These observations have revealed 
that deviations from a continuous radial or azimuthal distributions are common in protoplanetary disks. 
The most typical radial discontinuities are disk ``gaps", ``holes", and ``rings".  The term ``cavity" 
has also been frequently used to define a significant depletion of material occurring 
in the inner tens of AU, regardless of the presence of substantial material close to the star. In this 
section we define gap and ring as any azimuthally-symmetric deficit and enhancement in the disk brightness, 
respectively. In most cases (but not all), these deficits and enhancements are to be ascribed to a real 
depletion or concentration of material. Azimuthal discontinuities like lopsided rings or spirals are also 
often observed in disks, with the former structures being common in millimeter imaging and the latter in 
the visible/near-IR scattered light. 
We note that the disk regions with mass depletion are usually not completely devoid of material (i.e. gas is 
often observed inside dust gaps).
Gaps and holes observed with different tracers look different and do not always agree because individual 
gas and dust tracers have strong restrictions on the temperature, density, and grain size of the material they can detect. 
Some disks that show clear holes in mm interferometry show no holes in scattered-light imaging, and IR dust 
and gas observations find that these holes are not void regions but still host disk material. The reconciliation 
of these observations is currently a necessity, because while the incorrect use of definitions may produce 
confusion, the combination of different tracers bears the power to clarify the 
origin of disk gaps/holes and their link to disk evolution.

To date, it is still unclear how disk dispersal in the inside-out scenario may be linked to the rapidly 
increasing number of spatially-resolved observations of gaps and holes, spirals, and other radially 
or azimuthally asymmetric features detected at radii $> 10$ AU.
 The concept of disks that are ``primordial" and disks that are ``in transition"
may need to be revisited on the basis of new evidence from the growing number of spatially-resolved 
observations, which shows that disks may have structures previously attributed to evolution at phases much 
earlier than previously expected (e.g. ALMA image of HLTau).  

Clarifying the link between the disk gaps observed on small and large scales and the global disk structure is essential to 
understand disk evolution \citep[e.g.\,][]{Owen2016}. The growing number of observations that probe disk material 
within or beyond 10 AU starts now to provide grounds toward a unified picture of disk evolution and dispersal. For instance, 
ALMA can provide spatially-resolved images only down to disk radii $\gtrsim$ 3 AU 
at 120-140 pc (or $\gtrsim$ 30 AU in star-forming regions at 1 kpc)\footnote{An exception is the disk 
of TWHya in Andrews et al. 2016, where structures down to $\approx$ 1 AU are visible in the ALMA image thanks to the unique 
vicinity to Earth of this disk (54 pc).}. ALMA is also not optimal to observe disk gas at $\lesssim10$ AU due to a 
combination of angular and spectral sensitivity and the fact that the hot gas emits strongly in the IR, but not at mm 
wavelengths. On the other hand, rovibrational CO emission at NIR wavelengths is a good tracer of disk structure and 
gaps at 0.05-20 AU (Salyk et al. 2011, Banzatti \& Pontoppidan 2015), and NIR dust emission probes a similar region 
(see Section \ref{innerdisk-sect}), providing overlap and complementarity to the disk region probed by mm interferometers.
A global understanding of gaps therefore requires a combination of observations at different wavelengths 
to probe disk radii from the smallest to the largest distances from the star in both gas and dust.

In this section, we explore how different imaging techniques see gaps and asymmetries in disks (specifically mm 
interferometry and optical/IR scattered light imaging), and describe their limits, ranges, and degeneracies.

\subsection{Millimeter continuum interferometry observations}
Whereas dust gaps in disks were traditionally identified through a dip in the MIR part of their SED due to the 
deficit of warm dust \citep[e.g.][]{strom89,Forrest2004,Calvet2005,Brown2007}, their presence was confirmed 
through (sub)millimeter interferometric imaging at subarcsecond resolution,
using e.g. the SubMillimeter Array (SMA), Plateau de Bure Interferometer (PdBI) and Combined Array for Research in Millimeter-wave 
Astronomy (CARMA). The millimeter continuum images revealed that dust was indeed depleted from the inner tens of AU 
of the disk, showing ring-like outer disk structures (e.g. 
\citealt{Pietu2006,dutrey08,hughes09,Brown2009,Isella2010mwc758,Andrews2011}; see review in 
\citealt{WilliamsCieza2011}). 
Interestingly, some large mm-dust cavities were found in disks without a clear deficit in 
their SED, e.g. MWC758, UX Tau A and WSB 60, possibly due to vertical structure and small/large dust grain segregation \citep{Andrews2011}.

The image quality of these pioneering interferometers was rather low, due to the small number of antennas, resulting in low 
\emph{u,v}-coverage and S/N (typically 10-20 peak S/N ratios). The image results from the Fourier transform of the observed 
visibilities, using a deconvolution algorithm ('cleaning') to suppress the side lobes. 
Interpretation of these images thus has to be done with care, as the deconvolution process generally does not result in a 
beam-convolved image, but rather the best attempt of the algorithm to deconvolve the data with incomplete \emph{u,v}-sampling. 
The visibility data can be represented in a real and an imaginary component as function of baseline ({\emph{u,v}-distance), 
usually deprojected along the position angle and inclination of the disk \citep{Berger2007}. The real component represents 
the radial variations, and for a ring-like structure it shows an oscillation pattern (following a Bessel function), where 
the first 'null' is a measure of the cavity size \citep{Hughes2007}. Emission in the imaginary component indicates azimuthal 
asymmetries along the ring: zero emission indicates an axisymmetric disk. The real and imaginary components are measured with respect to 
the phase center, the center of the disk, so an offset will result in non-zero imaginary emission.

Interpretation of the dust continuum images is usually done by fitting the visibilities (in the \emph{u,v}-plane) with a 
radiative transfer model. Typical dust models include a dust surface density profile $\Sigma(r)$, following a power-law with 
or without exponential tail, and an inner cut-off at the dust cavity radius. This cut-off is usually taken to be sharp, to 
simplify the fitting and limit the parameter space, although this is generally considered to be unphysical\footnote{Higher spatial 
resolution observations are required to distinguish between sharp and smooth cut-offs \citep[e.g.][]{Andrews2011_Lkca15}.}.
When near infrared excess is measured in the SED, an inner disk is often added by setting $\Sigma(r)$ to non-zero between 
the dust sublimation radius and an arbitrary inner disk size (typically 1-10 AU). The inner disk may cast shadows on 
the gap edge, so this is a crucial part of the interpretation of mm data. 
As azimuthal asymmetries were usually not significant, a simple assumption of an axisymmetric disk was used, with zero imaginary emission.

The huge increase of sensitivity and \emph{u,v}-coverage by ALMA has resulted in many high quality images of disk dust gaps at 0.2-0.3" 
resolution \citep[e.g.][]{vanDishoeck2015}. The high S/N (typically $>$100) leaves no doubt about the azimuthally asymmetric nature of 
some of these disks. The most extreme examples are Oph~IRS~48 \citep{vanderMarel2013} and HD~142527 \citep{Casassus2013,Fukagawa2013} with 
contrasts $\sim$30--130. Minor azimuthal asymmetries with contrasts of $\lesssim$2 appear in SR~21 and HD~135344B 
\citep{Perez2014,Pinilla2015beta}. Several clearly axisymmetric dust rings are also found
\citep[e.g.][]{Zhang2014,Walsh2014,vanderMarel2015-12co,vanderMarel2016-isot,Canovas2016}. These structures can be understood in the 
context of mm-dust trapping in gas pressure bumps \citep[e.g.][]{Whipple1972,Pinilla2012b}. This is supported by the segregation of small 
dust grains as seen in scattered light, where observations reveal no or smaller gaps \citep[e.g.][]{garufi2013}, and the presence of gas inside the 
dust cavity as shown by ALMA CO observations \citep[e.g.][]{pont08,Bruderer2014,Zhang2014,SPerez2015,BP15,vanderMarel2015-12co,vanderMarel2016-isot,Canovas2016}. 
CO intensity maps (integrated over velocity) can resolve the gas cavities directly at $\sim$0.25" resolution, when 
their inner radius is large enough (several tens of AU). A quantitative analysis of these data indicates deep gas gaps, with  
density drops of several orders of magnitude, which are smaller than the dust gaps (van der Marel et al. 2016a). However, the 
amount of gas inside $\sim$10 AU remains unconstrained, as the emission at this resolution is dominated by the 
the edge of the gap. NIR observations (Section \ref{innerdisk-sect}) provide 
more information about the presence of gas closer to the star. The dust asymmetries may result from azimuthal trapping in a 
vortex, as a result of Rossby Wave instability in the pressure bump \citep[e.g.][]{BargeSommeria1995,Birnstiel2013,LyraLin2013}.

Considering the large parameter space and the high S/N, intensity profiles rather than full radiative transfer models are 
often used to fit these data, especially in azimuthally asymmetries \citep[][]{vanderMarel2013,Perez2014,Walsh2014,Pinilla2015beta,vanderMarel2015vla}.
 The edges are generally more consistent with a smooth ring (following a radial Gaussian) rather than a sharp cut-off \citep{Andrews2011_Lkca15}. 
Observing the continuum at different wavelengths reveals a wavelength dependency of the cavity size through the shift 
of the null in the visibilities \citep[e.g.][]{Pinilla2015beta,vanderMarel2015-12co} or as radial dependence of the spectral 
index $\alpha$, with  $F_{mm}\sim\nu^{\alpha}$ \citep[e.g.][]{Wright2015,Casassus2015,vanderMarel2015vla}, indicating that 
the larger dust grains are usually more concentrated.

As ALMA is reaching its full capacity, milli-arcsecond observations have revealed a possibly different type of gaps in 
disks: series of narrow bright and dark rings in the dust continuum of HL Tau and TW Hya \citep{ALMA2015,Andrews2016}, 
which are interpreted through a range of possibilities in the context of planet gaps, snowlines, magnetized disks, 
dust opacity effects, and sintering-induced dust rings \citep[e.g.][]{Dong2015,zhang2015,banz15b,Flock2015,Pinte2016,okuzumi2016}.

\subsection{Scattered light observations \label{scattered-sect}}

Scattered light observations in the visible and NIR probe the dust in the surface layers of the disk. At those wavelengths, 
a relatively small column density of dust is enough to attain a scattering optical depth of the order of unity. These observations mostly 
trace (sub-)micron sized grains, which are the dominant population of dust at the disk surface, as the strong coupling between small dust 
grains and gas attenuates their settling towards the midplane. 
To spatially resolve a disk in scattered light, high-contrast and high-resolution observations are needed. This has biased the 
sample of detected protoplanetary disks ($\sim 30$) towards bright disks around HAeBe stars in the nearest star forming regions 
\citep[see][]{quanz2015}. The primary limit on the stellar brightness is dictated by the adaptive optics system, whereas large disk 
brightnesses are needed to obtain a relatively high contrast with the stellar luminosity.  

Differential imaging techniques are used to overcome the large star-disk flux contrast at small angular separations. Post-processing PSF subtraction initiated the high-contrast imaging by means of coronagraphic \emph{Hubble Space Telescope} (HST) observations \citep[e.g.][]{grady1999,weinberger1999}. This technique is powerful to resolve the outer disk region, but limits the access to the inner $\sim 1''$ because of the limited telescope size and the need for a coronagraphic mask. Ground-based facilities like the VLT and Gemini provide dedicated differential imaging instrumentation \citep[e.g.][]{beuzit2006,macintosh2008}. Angular differential imaging (ADI) was developed for direct detection of companions, but it can be used for scattered light observations of disks. The principle of ADI is to keep the orientation of the telescope pupil fixed on the detector such that the field of view rotates around the target star. In this way, the disk signal rotates with respect to the quasi-static speckles and a reference PSF can be constructed from the target star itself \citep{marois2006}. This technique is particularly powerful for radially narrow disks, such as debris and edge-on disks \citep{milli2012} but it may suffer from flux losses by self-subtraction in extended disks \citep{garufi2016}. Polarimetric differential imaging (PDI) makes use of the polarizing nature of dust grains by taking the difference of orthogonally polarized images which subtracts the unpolarized stellar halo and speckles \citep[e.g.]{canovas2011,avenhaus2014}. Pioneering works were done by \citet{Kuhn2001} and \citet{Apai2004}, while the first systematic census of protoplanetary disks in PDI was performed with Subaru/HiCIAO  by the SEEDS consortium \citep[e.g.][]{Hashimoto2011, Kusakabe2012, Grady2013}.

The surface brightness of a disk in scattered light depends both on the disk structure and the scattering properties of the dust grains in the disk surface. For example, a local change in surface density or pressure scale height will affect the irradiation of the disk surface and the amount of light scattered into our line of sight. For inclined disks, the surface brightness is also determined by the dust properties because of the scattering angle dependence on the phase function and the degree of polarization. Scattered light provides also insight into the dust properties in the disk surface through measurements of disk color \citep[e.g.][]{mulders2013,stolker2016} and phase function (Stolker et al. subm.).
While small (compared to the observed wavelength) grains scatter isotropically, the phase function of larger grains has a forward scattering peak which can manifest itself as a brightness asymmetry of the near and far side of a disk \citep[e.g][]{mishchenko2000,thalmann2014}. On the other hand, the degree of polarization typically peaks around scattering angles of 90 degrees, which is near the disk major axis \citep[e.g.][]{hashimoto2012,min2016}. The combined effect of disk structure, phase function and degree of polarization can make the interpretation of polarized surface brightness non-trivial: Disentangling the different effects may require radiative transfer modelling. Here, sub-millimeter observations can help to trace a complete picture of the distribution of small dust, large dust, and gas throughout a disk (see Section \ref{discussion-sect}).

Several types of morphological features and brightness asymmetries have been detected in scattered light \citep[e.g.][]{casassus2016}. 
Spiral arms have been observed in a number of transition disks \citep[e.g.][]{muto2012,garufi2013,wagner2015}. Their origins are still 
debated due to our limited knowledge of their vertical structure, since in principle both global changes of the dust properties and 
small variations on the disk scale height may account for the observations. The observed spirals
can be produced by various mechanisms, including planet/stellar interactions with the disk \citep[e.g.][]{ogilvie2002,boss2006}, 
gravitational instabilities \citep[e.g.][]{cossins2009}, and shadowing effects \citep{montesinos2016}. The visibility of a spiral 
density wave in scattered light depends on the strength of the temperature and/or surface density perturbation \citep{juhasz15}. 
A massive planet can trigger both a primary and secondary spiral arms interior to its orbit \citet{dong2015}, which 
resembles some of the observed spiral arms \citep[e.g.][]{Benisty2015}.

Brightness asymmetries may also be related to global or local asymmetries in the disk structure. In some cases a 
plausible connection between disk surface and midplane can be made when the asymmetry in scattered light and sub-mm dust 
continuum coincides \citep[e.g.][]{garufi2013,marino2015b}. 

Radial reductions in surface brightness are often interpreted as gaps \citep[e.g.][]{quanz2013,thalmann2015,rapson2015}.
Nevertheless, a decrease in the scattered light flux does not necessarily relate to a decrease in gas and/or dust surface 
density but could also be a shadowing effect \citep[e.g.][]{siebenmorgen2012,garufi2014}.
Local shadowing effects have been detected on a few disks: for example, a warped inner disk \citep{marino2015a}
can produce azimuthal surface brightness reductions, possibly variable on detectable timescales \citep{pinilla2015,stolker2016}. 
 Radiative transfer modelling, ideally combined with hydrodynamical simulations, are required to translate scattered light 
flux into gap depth \citep[e.g][]{fung2014,rosotti2016}. In gaps opened by planet formation \citep[e.g.][]{baruteau2014}, 
the gap depth depends on the planet-to-star mass ratio, the disk aspect ratio, and the turbulence \citep[e.g.][]{kanagawa2015}. 
Alternative explanations include the effect of snow-lines on the dust surface density \citep[e.g.][]{zhang2015,banz15b,okuzumi2016}, 
dust evolution \citep{birnstiel2015} and 
vortices at dead zone edges \citep[e.g.][]{varniere2006}. Non-axisymmetric gap edges in scattered light can be shaped by 
dynamical disruption by a planet \citep[e.g.][]{casassus2012}, but they can also be an illumination effect of an inclined gap 
edge \citep[e.g][]{thalmann2010} or a shadowing effect by a misaligned inner disk \citep[e.g.][]{thalmann2015}.

Finally,  the detectable size in scattered light is limited by the disk
structure and the sensitivity of the instrument. For example,
the $\tau$=1 height of a flaring disk will increase with radius as long as the surface
density is high enough. This means that, depending on the disk structure,
the disk becomes self-shadowed  at a given radius and what we observe in scattered light
beyond that radius is an optically thin/faint surface layer. Moreover, the illumination by the star
decreases as r$^{-2}$, so that disks are not detectable any more in scattered light beyond a certain
radius.
For PDI scattered light images, the
sensitivity rapidly drops down at 1-2 arcsec, 
whereas  observed disks are often larger. On the other hand, HST coronographic scattered light images work better above 2-3 arcsec \citep{grady05},
although this cannot be applied to disks with sizes $<$200-300 AU.
The same applies to the very inner part of 
the disk (at the dust sublimation radius), unachievable with current instrumentation. Very compact 
disks also remain unresolved with mm imaging and undetectable in scattered light 
\citep[e.g.][]{garufi2014}. If there is a class of disks with $<$10-30 AU radii \citep[e.g.][]{woitke13}, then SPHERE 
and ALMA would be the right instrument to measure their outer edge.


\section{The disk mass \label{diskmass-sect}}

In this section we address disk mass estimates from different observations and how they can be compared.
In particular, we address the issues of disk mass estimated from dust vs from gas, including the degeneracies due to assumptions 
on the dust sizes/properties/distributions, disk temperature, and the gas/dust ratio, and their implications for our understanding of disks.

\subsection{The dust mass}

The total mass of disk-forming material is distributed between the refractory dust (${\sim}1$\% by mass) and the volatile gas (remaining $\sim$99\%). 
Given the relative ease of broadband continuum observations as compared to spectrally resolved observations 
of atomic and molecular transitions, dust masses are generally easier to estimate. The total disk mass is 
then derived scaling up the dust mass by a standard gas-to-dust ratio, $\Delta_{\rm g/d}=100$. This 
standard ratio is being increasingly questioned, as the processes happening in disks 
(photoevaporation, planet formation, viscous evolution) are expected to affect the gas/dust ratio,
including radial variations, now clearly exposed by the differences in dust-gaps and gas-gaps (Section \ref{gaps-sect}). 
The mass of dust can be estimated from a continuum measurement and an 
adopted dust opacity, by assuming a mass-averaged dust temperature, often in the $20$--$30\,$K range for TTS disks 
and somewhat higher for HAeBe disks \citep{andrews13}.
The grain size distribution, porosity, and composition affect the dust opacity, considering that disk and ISM dust can 
be very different. This in turn 
affects the thermo-chemical models of the disks. 

Dust grains are poor emitters of blackbody radiation above 
wavelengths $\sim 2 \pi \times$ radius, so long wavelengths can be used to examine large particles to discern 
what systems look promising for planets. Finding radio emission with a dust-like spectral index is thus a 
clue to the presence of grains up to centimeter-sizes. This field of study was pioneered by 
\citet{wilneretal2005}, who detected dust at 3.5 cm in TW Hya, and similarly large grains have since 
then been found in many
other objects \citep{rodmann06,ricci10}. Advances in sensitivity with instruments such as VLA and MERLIN make 
it possible to image cm-sized grains even at very low surface brightnesses. 
In the future, this science will be opened up with the Square Kilometre Array, especially in later phases when $\sim 1000$km baselines at several-GHz frequencies 
could be used to obtain few-AU resolution at the distances of nearby star-forming regions. 

For typical grain size distributions in disks, it is usually true that the mass is mostly in the large particles, while the emission comes mainly 
from the smaller grains (due to a more favourable surface-area to mass ratio). However, if extended up to 
the size of planetesimals, this means that most of the disk mass is unobservable 
by radiation signatures. Hence, where $M_{disk}$ is deduced from data, it should strictly 
refer to a size of particles contributing significantly to the emission at the wavelength 
of observation. Where measured dust disk masses are below those required to build 
planetary cores, it may be that planet formation has already occurred, and what we observe 
is remnant material \citep{greaves+rice2011}.

Very low-mass, anemic, or dust-depleted disks \citep{lada06,currie09} have low submillimeter and millimeter
fluxes, which makes it
hard to detect them at long wavelengths. In addition to evolved disks, other disks are intrinsically faint, such as disks around brown 
dwarfs. For these cases,
the mid- and far-IR data may be a good option to set strong constraints to the total dust
content \citep{currie11,harvey12, sicilia11,sicilia13a,sicilia15a,scholz16}, even though 
the degeneracy between disk scale height and total disk mass cannot be broken unless long-wavelength
data is included.

\subsection{The gas mass \label{gasmass-sect}}

Although molecular hydrogen is the main component of the disk mass, the total H$_{2}$ mass cannot be directly 
measured. H$_{2}$ does not emit at temperatures found in cold disk regions ($T_{\rm gas}{<}100\,$K) due to 
its lack of a permanent dipole moment, and alternatives such as vibrationally excited H$_{2}$ gas trace only the 
hottest part of the disk. Moreover, H$_2$ is optically thick in the parts of the disk where emission originates,
not even allowing a regional mass determination \citep{carmona08,bitner08}. 
Line-of-sight absorption is another interesting, but poorly explored, method to trace 
a part of the H$_{2}$ mass, but difficult to apply in practice \citep[e.g.][]{france14,martinzaidi08}. 
 This has led to great interest in alternative tracers of the gas mass, the main of which we review below.

{\bf Carbon monoxide (CO): }The most commonly used cold gas tracer is rotational emission from the CO molecule,
 which in disks has an abundance of $^{12}$CO/H$_{2}{\approx}10^{-4}$ \citep{thi01,france14}. The 
$J_{\rm upper}{=}1$, $2$, and $3$ transitions, as well as several higher-lying ones, are observable 
from the ground with, for example, the ALMA, NOEMA, and SMA interferometers and the APEX, ASTE, and 
IRAM~$30$-meter telescopes. Extensive archival data exist for JCMT, CSO, and \emph{Herschel}. 
Optical depth effects can be corrected for by observing the less abundant isotopologs $^{13}$CO, C$^{18}$O, 
and C$^{17}$O. Modelling of these needs to include isotopolog-selective (photo)chemistry 
(Miotello et al. submitted; \citealt{miotello14}). Simpler models calibrated with detailed 
simulations can also yield good estimates of the gas mass, although with no provision for global carbon 
depletion \citep{williams14,kama16a,kama16b}. CO-based gas masses are often a factor of $10$--$1000$ lower than 
expected from the interstellar standard gas-to-dust ratio of 100 \citep{dutrey97,thi01}. These low CO fluxes 
may signal the depletion of carbon and oxygen from the warm, UV-irradiated gas of the disk surface layers, 
and highlight the need for complementary tracers of the total gas mass 
\citep{bruderer12, favre13, du15, kama16b}.

{\bf Hydrogen deuteride (HD):} The singly deuterated isotopolog of H$_{2}$ is a powerful probe of the total 
warm gas mass in a disk. The two lowest rotational lines of HD are at $112$ and $56\,\mu$m, and require 
space-based observations because of the high atmospheric opacity. The first and to-date only published 
detection of HD was obtained towards TW~Hya by \citet{bergin13}, who found a total disk mass of 
${\approx}0.05\,$M$_{\odot}$. \citet{mcclure16} expand the sample of 
$3\sigma$ detections with DM~Tau ($4.5\times10^{-2}\,$M$_{\odot}$) and GM~Aur ($19.5\times10^{-2}\,
$M$_{\odot}$). The upper limits on HD lines towards HD~100546 constrain the gas-to-dust ratio in that system 
to ${\leq}300$, equivalent to a gas mass of $\leq 2.4\times 10^{-1}\,$M$_{\odot}$ \citep{kama16b}.

{\bf Atomic oxygen ([OI]): }Neutral atomic oxygen traces warm-to-hot gas in the disk atmosphere, but 
the analysis can be thwarted by contamination issues. For late-type stars and disks with no residual 
envelope, the far-infrared $63$ and $145\,\mu$m lines of [OI] are in principle a clean probe of the 
warm oxygen or even total gas mass \citep{woitke10, kamp11}. However, this assumes a 
standard total gas-phase oxygen abundance. Depletion of volatile oxygen from the disk atmosphere by 
sequestration into planetesimals forming in the midplane can reduce the oxygen abundance globally by 
several orders of magnitude, making it impossible to derive the gas mass from [OI] alone 
\citep{du15, kama16b}. Many transitional TTS disks have [OI] fluxes approximately a 
factor of two lower than ``full" or ``primordial'' disks with the same far-infrared continuum luminosity, 
while the transitional disk [OI] flux range also contains all ``primordial'' disks \citep{keane14}. 
The cause of this is not yet clear, but a low gas-to-dust ratio or overall oxygen depletion are 
potential explanations. For early-type stars, where the disks are warmer and depletion of volatiles 
may be less important, the [OI] flux sometimes carries a non-disk contribution \citep{dent13} and 
gives an upper limit on the warm gas mass. This is underlined by the case of HD~100546, where a 
circum-disk envelope adds to the $63\,\mu$m line flux \citep{bruderer12, kama16b}.

\begin{figure*}
\begin{center}
\includegraphics[width=17cm]{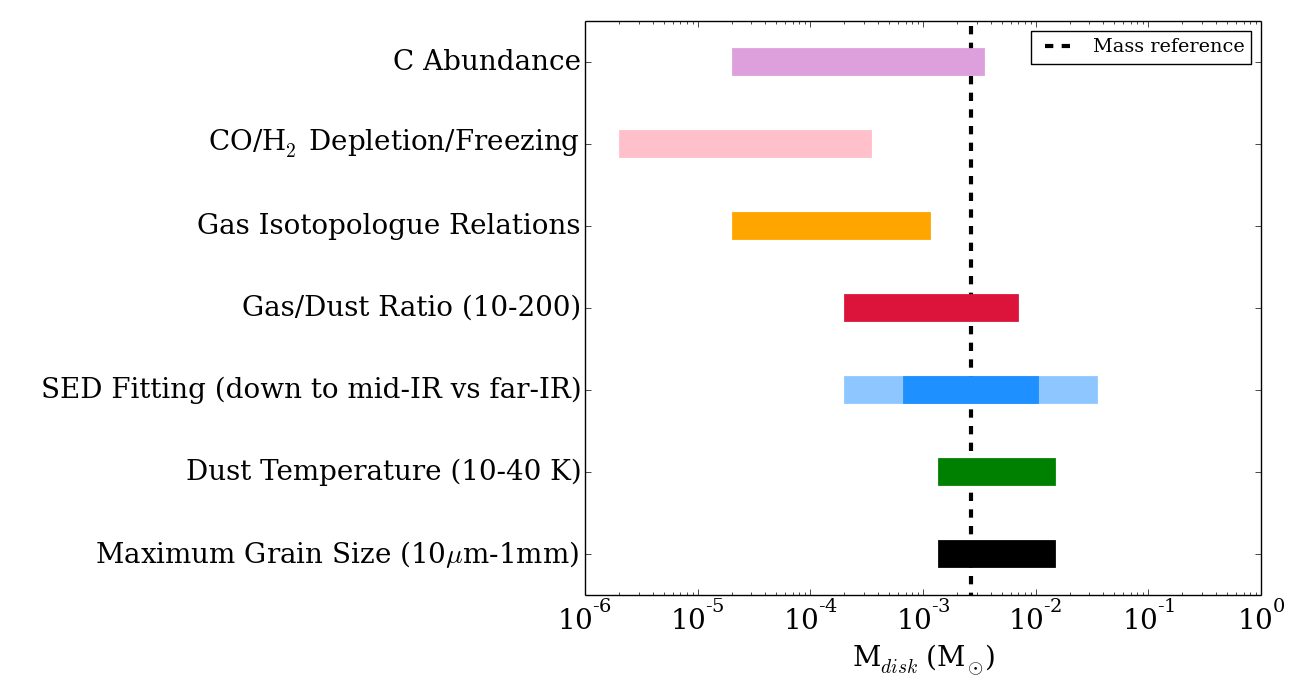}
\caption{Ranges of disk masses resulting from the uncertainties in total masses derived from dust and gas and measured by various techniques.
The dashed vertical line indicates our reference mass for this exercise, taken to be the mass estimate for a disk with
1.3mm emission of 25 mJy around a 1M$_\odot$ star \citep[the median value in][]{andrews13}.
The colored bars mark how the mass estimate may change depending on the
method. Using a gas tracer, depending on: C abundance \citep[purple;][]{kama16a}, CO depletion/freeze out 
\citep[pink,][]{thi01,du15}, isotopologue relations \citep[yellow;][]{miotello14}, and changing
the gas/dust ratio between 10-200 \citep[red;][]{panic09,riviere13}. Using a dust indicator: with a complete SED lacking the mm
data but including mid-IR (light blue) and far-IR \citep[dark blue;][]{sicilia11,sicilia15,currie11}, 
varying the assumed dust temperature \citep[green;][]{andrews13}, and changing
the maximum grain size between 10$\mu$m and 1mm \citep[black;][]{miyake93,henning96}.}
 \label{massunc-fig}
\end{center}
\end{figure*}

\vskip 0.4truecm

Figure \ref{massunc-fig} offers a visualization of the various uncertainties to which different disk mass tracers are subject. For this exercise, we
take as a reference the mass derived for the disk around a 1M$_\odot$ star at 140 pc distance with a 1.3mm flux of 25mJy, following the methods 
in \citet{andrews13}. This is roughly the mean value for solar-type stars in \citet{andrews13}. After deriving this dust-based total disk mass, we
made the experiment of considering it as the ``true mass" of the disk, 
and calculate the mass values that other different methods would measure, based on their own uncertainties\footnote{Note that for any other 
initial ``true mass", all the ranges would simply shift as a bulk to higher or lower masses.}. 
For dust-based estimates,
we explored the effect of grain growth \citep{miyake93,henning96,henning10}, dust temperature \citep{andrews13},  and of the gas/dust ratio \citep{riviere13,panic09}.
As for gas-based measurements, we explored gas depletion
\citep{thi01,du15,kama16a,miotello14} for several species and the mass values we would obtain 
accounting for the typical gas-phase C depletion, including freezing-out\citep{thi01}, carbon depletion \citep{kama16a}, and gas isotopologue
relations \citep{miotello14}. The uncertainties in
case the disk mass is estimated from an incomplete SED (lacking mm data) are also shown, 
as they are important to estimate the mass dispersal timescales,
including low-mass and evolved disks (too faint for most mm-wavelength instrumentation). From this figure, the intrinsic uncertainties in
all estimates of disk masses are revealed, as well as the importance of finding reliable gas
tracers to estimate reliable disk mases is the main open problem regarding disk masses. 

The possibility of high-sensitivity gas tracers with ALMA would
be a key, both to obtain better mass estimates as well as to resolve the potential radial dependency of 
grain sizes and gas/dust ratios throughout the disk.
A robust way to estimate gas masses would require simultaneous modeling of spatially-resolved
CO isotopologue data, together with sub-mm imaging, and also [O I] 63$\mu$m and [C I] emission
\citep{woitke16,kama16a,kama16b}. Building on these and other works, future ALMA observations together
with a better understanding of the disk chemistry will be keys to determine the
disk masses.


\section{Accretion \label{accretion-sect}}

In this section we discuss the observables of mass accretion onto the star, and 
several outstanding problems raised by observations.
Accretion plays a central role in disk dispersal: Angular momentum transport
and energy minimization in the disk, driven by viscosity, cause mass transport inwards and accretion
onto the star, while it also produces expansion of the disk outer radius in time \citep{gammie96,hartmannetal98,hartmann06}. 
While viscous evolution alone would require disk
evolutionary times much longer than observed \citep{hartmannetal98}, accretion is a powerful mechanism:
it connects the whole disk and the star, having the potential to affect the early stellar
evolution \citep{baraffe09}, the architecture of the nascent planetary system and the migration of planets \citep{lubow10}, 
and the disk structure (such as the mass distribution in the 
inner disk and the dust vs gas disk radius). 

There are two main theories to explain how accretion proceeds onto the star: boundary layer 
(BL; the gas accretes directly from the disk to the central star) and magnetospheric accretion (MA; 
the gas from the inner disk channeled through the stellar magnetic field lines). 
Early works focused on the BL scenario for both TTS and HAeBes 
\citep{Bertout88,BasriBertout89,Blondel94,Blondel06}.
Nevertheless, MA is since long widely accepted for TTS \citep{Uchida85,koenigl91,Shu94,Alencar07}, 
supported by evidence from near-UV excess, emission line profiles, 
observed magnetic (B-) fields, rotational modulation of line profiles, and the presence of outflows and jets. 
MA also seems to be drive accretion in brown dwafs \citep{Riaz13} and has been temptatively proposed 
for accreting planets in formation \citep{Lovelace11,Zhu15}. Several lines of evidence suggest that accretion could 
also be magnetically driven in late type HAe stars, as suggested by spectro-polarimetry \citep{Vink02,Mottram07} and near-UV continuum 
excesses \citep{Muzerolle04,DonBrit11,Mendi11b,Mendi13,Fairlamb15}. Near-UV/optical/NIR spectral lines also show
profiles similar to those of TTS \citep{Mendi11a,Cauley14,Cauley15}, which can be reproduced from MA line 
modelling (e.g. UX Ori and BF Ori in \citealt{Muzerolle04} and \citealt{Mendi11b}, respectively).

The small/non-detected B-fields in HAeBes is commonly argued against MA operating in 
these objects. Although their internal structure with radiative envelopes did not predict the presence of
B-fields and related high-energy emission, X-rays are regularly detected toward HAeBe stars 
\citep[especially, among those with M$<$3M$_\odot$, e.g.][]{feigelson03,preibisch05,forbrich07, stelzer09}. 
The minimum B-field required to drive MA is strongly dependent on the stellar properties \citep{Krull99}, 
so if B-fields of $\sim$ 1 $kG$ are necessary in TTs, much smaller B-fields of only hundreds of $G$ or less would be enough for 
the HAeBes \citep[see the discussion in][]{Mendi15b,Fairlamb15}. There are clear indications that the accretion 
mechanism changes at some point within the HAeBe regime \citep[e.g.][]{Mottram07,Mendi11b,Fairlamb15}, 
and there are several early-type HBes for which MA is definitely not able to reproduce the strong near-UV excesses
observed \citep{Mendi11b,Fairlamb15}. Understanding accretion in
HBes would need a new approach, perhaps returning to BL \citep{Cauley15} or 
considering similar mechanisms (but for accretion, not decretion) as in classical Bes \citep{Patel15}.

\begin{figure*}
\begin{center}
\includegraphics[width=15cm]{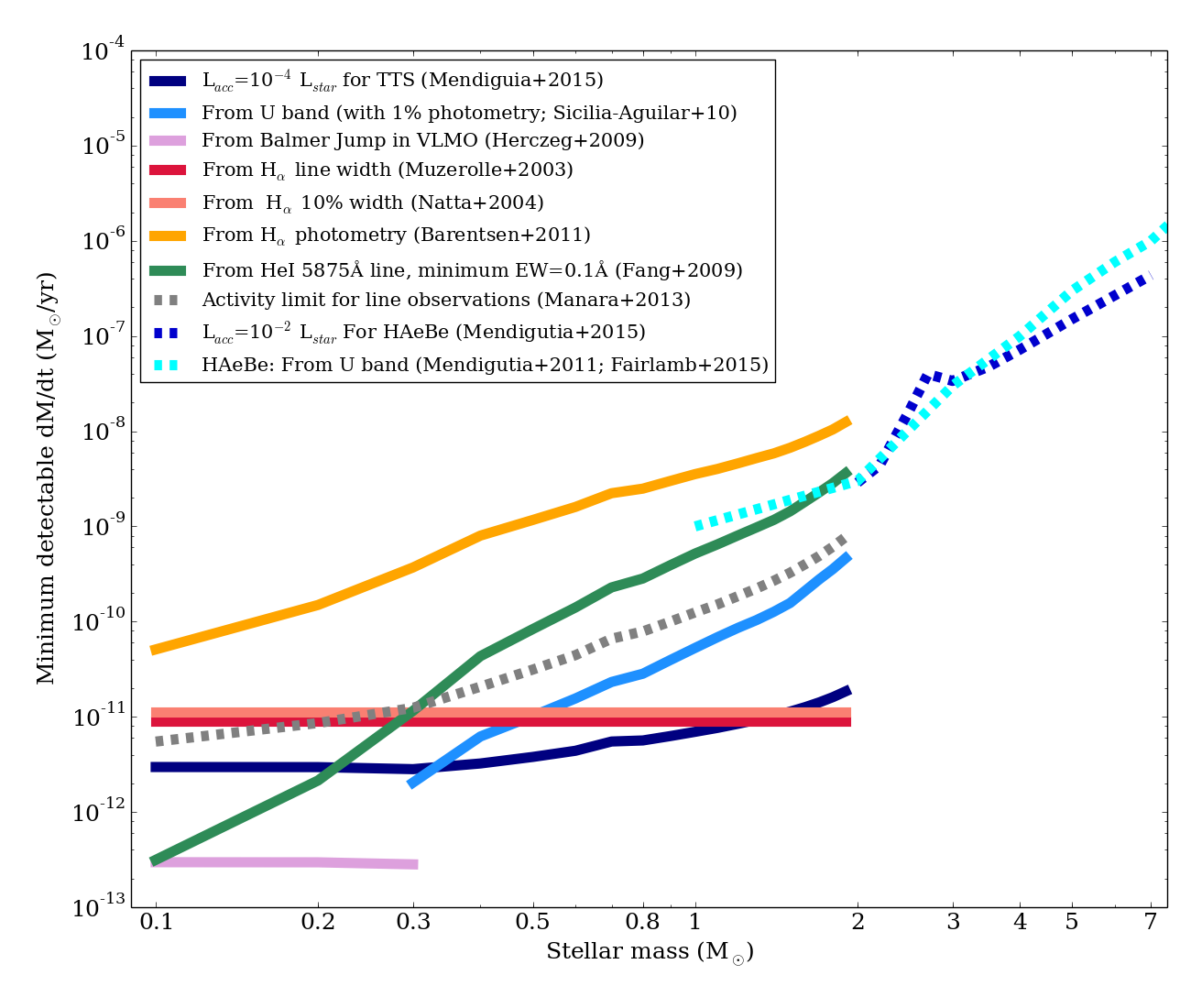}
\caption{Lowest detectable accretion rates for stars with different masses, using different techniques.
The \citep{siess00} isochrone track for 3 Myr-old stars is used to transform between mass and luminosity.
A distance of 140 pc is assumed. See references in text for details on the various techniques.}
 \label{accretion-fig}
\end{center}
\end{figure*}

Observationally, accretion rates are ultimately derived from the accretion luminosity. The accretion luminosity can
be estimated from the veiling, continuum excess \citep{gullbring98}, or emission line luminosity
\citep{natta05,fang09,alcala14}. To determine the luminosity due to accretion, the spectral type and extinction of the
star must be well-constrained, especially if the
accretion luminosity is small compared to the stellar luminosity. 
The lowest accretion rate that can
be detected depends on how other processes, such as activity and winds, affect the accretion
tracers \citep{sicilia10,manara13}. This makes the methods relying on direct accretion luminosity estimates
(measuring the veiling, the Balmer jump, or U band excess) more powerful than those
relying on line emission (as showed in Figure \ref{accretion-fig}). 
When detailed information on the stellar properties (spectral type, luminosity, extinction,
typical activity levels in similar but diskless stars) is
available, the detection limits for accretion onto solar-type stars (late K-early M) can be
as low as 10$^{-11}$ M$_\odot$/yr \citep{sicilia10}, while for very low-mass
stars and brown dwarfs (BD), accretion rates as low as 10$^{-13}$ M$_\odot$/yr can be inferred \citep{natta04,herczeg09}.
On the other hand, accretion-related spectral lines have the advantage of providing velocity information.
Metallic lines 
observed in accreting stars \citep{hamann92} span a large range of critical densities
and temperatures, thus tracing material in various physical conditions and different locations within the
accretion-related structures.
The velocities can be used to estimate the extent of accretion columns via Doppler tomography, 
using the strong H$\alpha$ and H$\beta$
lines \citep{muzerolle01,muzerolle03,lima10,alencar12}, or the many metallic emission and absorption
lines \citep{beristain98,beristain01,sicilia12,petrov14,sicilia15b}. 

\subsection{Accretion as a probe of the disk and the star}

Accretion involves the whole disk, as transport through the disk is needed to maintain the accretion
rate over time. 
Therefore, the presence (or lack) of accretion is a powerful tool to investigate disk structure and
physical processes. Although it is possible to measure accretion rates down to 10$^{-11}$ M$_\odot$/yr for solar-type
stars\footnote{Very low-mass stars and BD have typically lower accretion rates.}, we find that only very few stars have such low rates, suggesting
that accretion stops quickly after reaching levels below 
$\sim$10$^{-10}$ M$_\odot$/yr \citep{sicilia10}. This is consistent with the predictions of photoevaporation 
as a process removing the inner gaseous disk in a relatively short time \citep{clarke01,alexander06,gorti09}.
For solar-type stars stopping accretion seems very
rare unless dramatic changes have occurred to the whole disk \citep[such as strong mass depletion][]{sicilia13b,sicilia15a}. This is
in agreement with accretion being a global process that connects the inner and the outer disk and can be used as 
an indicator of global disk evolution \citep{sicilia15a, manara16}.

The presence (or lack) of accretion in disks with inner holes (identified from SEDs) also clearly separates two classes among transition disks
around T Tauri stars: accreting and non-accreting ones. 
While essentially all primordial or ``full" disks show
signs of accretion \citep{sicilia13b}, between 30-50\% of transition disks show none 
\citep[narrow emission lines, no UV excess, no veiling;][]{fang09,sicilia08b,sicilia10}. 
Although less explored, differences in the accretion rate between
primordial and transitional disks have also been observed for HAeBe stars \citep{Mendi12}. 
Our understanding of the evolutionary stage of transition
disks will improve with multiwavelength data (see Section \ref{discussion-sect}).

Accretion is also connected to the stellar properties, since the accreting
matter is channelled onto the star by the stellar 
magnetic field  \citep{koenigl91}. Therefore, studies of the properties and distribution of accretion columns
can also help to study the
magnetic field topology of the star, intimately related to the stellar structure and evolution (e.g. \citealt{donati11,donati13,gregory14}).
Time-resolved H$\alpha$ and metallic line emission spectroscopy are promising ways to study the presence, distribution, and
evolution of accretion-related hot spots, which can be connected to the structure of the stellar 
magnetic field \citep{alencar12,sicilia15b}. Extension of these studies to other stars
have the power to provide new information on stellar properties during a key time in their evolution.

\subsection{Open problems for accretion: disk masses and accretion tracers}

\begin{figure*}
\begin{center}
\includegraphics[width=15cm]{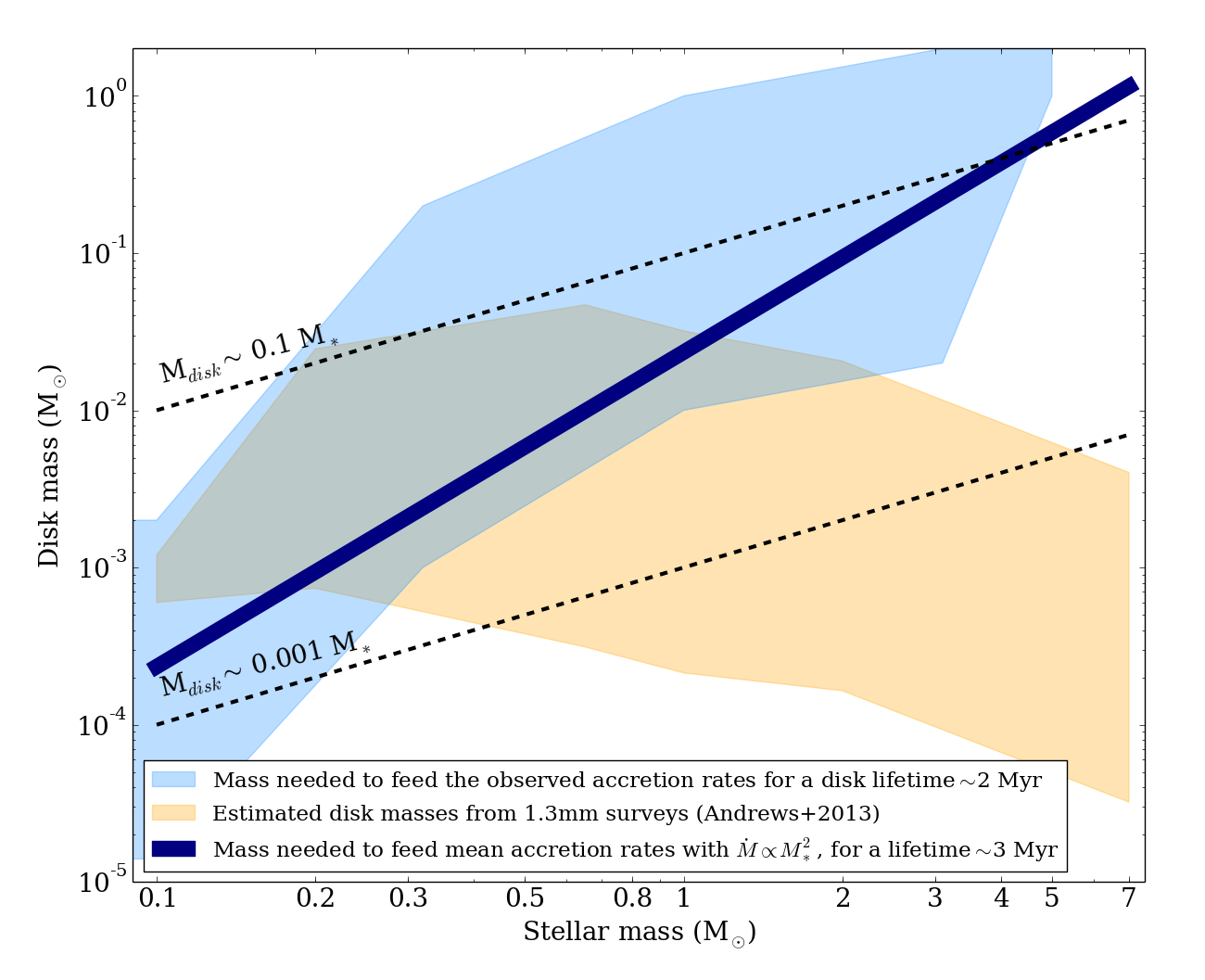}
\caption{Disk masses as measured by observations, and as expected from the need to support the observed accretion rates during a
lifetime of 2-3 Myr. The black dotted lines represent the usual
limits of disk masses between 0.1-10\% of the mass of the star. The yellow area displays the disk masses measured by
\citep{andrews13}. The blue region represents the expected disk masses for the whole range of
accretion rates observed, and a disk lifetime of 2 Myr (which is typically lower than the median disk lifetime of $\sim$3 Myr).
The dark blue line represents the expected disk masses considering the median accretion rate at 4 Myr and an accretion
lifetime of 3 Myr \citep{sicilia10}. }
 \label{diskmass-fig}
\end{center}
\end{figure*}

An outstanding problem regarding the disk's gas content is that accretion rates and observed disk masses are often in conflict:
considering the measured disk masses (dust- or gas-based) and the observed accretion
rates, a significant fraction of stars are expected to fully drain their disks in 
timescales shorter (sometimes, by several orders of magnitude) than the usual disk lifetimes. 
The situation can be extreme for some strong
accretors with not-so-massive disks  \citep{hartmann06,sicilia08a,sipos09,liu16,kospal16}.

Figure \ref{diskmass-fig}
shows the extent of the problem: if we compare the observed disk masses \citep{andrews13} with the mass ingested by
the star during 2 Myr (which is shorter than the median disk lifetime of $\sim$3 Myr; \citealt{sicilia06a}), there is 
a clear divergence. 
This is a problem that affects both TTS \citep{AndrewsWilliams07,sicilia11} and HAeBes \citep{Mendi12}, although
the difference is negligible for stellar masses M$_* <$0.2 M$_\odot$ and increases with the mass of the star (see Figure \ref{diskmass-fig}). 
Assuming that stars have variable accretion and that they only spend a small part of their lives accreting at very high
rates could help solving the problem, although there is no evidence of short-term strong accretion 
variability among most 1-10 Myr old stars \citep{sicilia10,costigan12,costigan14}.
Moreover, the problem still persits if the accretion rate is taken to be the median accretion rate observed for stars aged $\sim$3-4 Myr \citep{sicilia10},
scaling the results from solar-type stars to higher and lower masses following the usual \.{M}$\propto M_*^2$ relation \citep{natta05}.
This points to an overestimation of the accretion rates (or the time stars spend accreting), or an underestimation of the disk masses, or both.

At present, the uncertainty in the accretion rates (especially if derived from UV or accretion luminosities
for stars with well-known spectral types and extinctions) is smaller than the uncertainty
in the total disk mass (or at least, in the part of gas mass that we can measure; see Figure \ref{massunc-fig}).
Freezing-out of gas tracers and the potential changes in the gas/dust ratio as the disk 
evolves are clear problems in the estimation of gas masses.
We may be able to solve this mismatch as
more detailed, spatially-resolved gas observations become available.
The problem could be solbed by assuming gas/dust ratio $\sim$1000, but current observations, including data from
several Herschel Consortia, rather suggest that if any,
the gas/dust ratios may be lower than in the ISM \citep[e.g.][]{riviere13}.

Shorter disk and accretion lifetimes for HAeBe could contribute to solve the problem, and indeed 
accretion rates in HAeBe are expected to drop more abruptly than for TTS \citep{Mendi12}.
Changing the way the accreted matter reaches the star does not help: BL results in
higher accretion rates than MA, making the problem worse.
Considering that the correlations between the accretion rate and the disk 
mass expected from viscous accretion models \citep{hartmannetal98} often remain elusive,
the mass/accretion disagreement may be a signature of our difficulties estimating total disk masses, where dust masses are
by far better determined than gas masses -- but they only trace a minimal part of the disk \citep{manara16}. 

Another open problem is whether all accretion tracers measure the same thing.
The observed correlations between line and accretion 
luminosities do not necessarily indicate a physical relation between both \citep{Mendi15a}.
Spectro-interferometry of H$\alpha$, Br$\gamma$ and CO, with instruments on Keck-I, 
CHARA and VLTI, shows that although the Hydrogen lines are often used as a proxy to estimate
 mass accretion rates, the bulk of the emission arises in regions more extended than expected from
 magnetospheric accretion. Indeed, several observations reveal extended emission, consistent with 
the base of a disk wind \citep{kraus08,benisty10c,weigelt11,garcialopez15,caratti15}. In the same objects, in particular 
the most massive ones, the displacement of the photocenter is consistent with 
Keplerian motion \citep{ellerbroek14,kraus12,Mendi15b}, although it might be difficult to disentangle the emission from the 
disk from the emission from the base of the disk wind \citep{kurosawa16}. 

Moreover, observed emission lines often suggest that accretion-related structures are not monolithic entities,
but include gas with different physical conditions.
Several stars (DR Tau,  \citealt{petrov11}; S CrA,  \citealt{gahm08}; 
EX Lupi,  \citealt{sicilia12,sicilia15b}; RU Lupi,  \citealt{dodin12}) 
show evidence of line-dependent veiling or ``veiling-by-lines". 
In these cases, part of the accretion luminosity is not emitted as continuum, but as spectral lines, filling in
the stellar photospheric lines to different extents.
The strength of the veiling on a given line depends on the physical conditions on the accretion structures
(temperature, density) and the line formation conditions/atomic parameters of the given line.
The properties and structure of the accretion column may vary, depending on their number and structure, resulting in
strong line differences even in objects with similar accretion rates \citep{sicilia15b}.
Line-dependent veiling can dominate the spectrum during times of high accretion, resulting in strong, broad line emission \citep{kospal08,sicilia12,holoien14}. 
Line veiling has been poorly explored for
the general population, and could potentially affect the estimates of accretion rates, especially for objects with very
low accretion or in cases where the accretion columns are optically thin. 
Nevertheless, line-dependent veiling can be also used as a
tool to estimate the physical conditions in the accretion columns and accreting material
\citep{sicilia12,sicilia15b} (see Section \ref{timeresolved-sect}),
although extracting broad, faint lines from the stellar spectrum can be hard.


\section{Disk insights from variable phenomena and disk dynamics \label{timeresolved-sect}}

In this section we discuss the stellar variability and the dynamical effects observed in the disk as means to
explore the time dimension in disks. Different parts of the star+disk system respond to different timescales 
(stellar rotation, keplerian rotation), and the physical processes involved in disk dispersal are also
strongly dynamical.

\subsection{Photometric and spectroscopic variability}

Photometric and spectroscopic variability is a defining characteristic of pre-main sequence stars (PMS;
\citealt{joy45,herbst94,briceno01,alencar01}). Photometric monitoring campaigns at optical wavelengths 
covering timescales of days to months reveal three main causes for the 
variations in low-mass TTS \citep{herbst94}: periodic, cold (chromospheric) spots on the stellar surface, non-periodic 
hot (accretion-related) spots, and more extreme dimmings (up to 3-4 magnitudes in the optical on timescales of 
one week) accompanied by polarization changes, caused by circumstellar material in the line of 
sight \citep[UXOr-type variability; see e.g][]{Grinin91,natta97,Rodgers03}. 
A sensitive photometric survey with \emph{CoRoT} identified several 
common types of variability and was able to probe the presence of stable and unstable accretion, 
stellar hot spots, and common occultations by inner disk structures such as accretion 
columns and warps \citep{mcg15}. 
While hot and cold spots explain most of the variability observed in TTS, 
HAeBes, without cold spots, no chromospheres, and accretion shock temperatures comparable to their
effective temperatures, are dominated by UXOr-type variability \citep{Muzerolle04,Mendi11b,Mendi11a}.

The \emph{Spitzer Space Telescope}, especially during its warm mission, has been used to study infrared 
variability in a number of nearby clusters \citep{moralescalderon09,pop15,reb15,mor11,gun14,wol15,fla13,cod14}. 
These studies find that 60-90\% of the young 
stellar objects are variable in the infrared, with variability being stronger and more common among Class I 
sources and weaker and less common among Class III sources, with typical fluctuations of a few tenths 
of a magnitude. While many of these studies are focused on daily-to-weekly fluctuations, 
year-to-year fluctuations have been seen by both Spitzer \citep{reb14,meg12} and 
ground-based NIR studies \citep{sch12,wol13,par14,ric15}.
Full disks tend to show a flatter wavelength dependence of their variability over the 
2-10$\mu$m region \citep{meg12,kos12}, but some changes in color have been seen. The
3-5$\mu$m variability can be explained by a mix of extinction and disk 
variability \citep{pop15,wol15}, while ground-based NIR studies find similar results 
along with additional contributions from star spots \citep{car01,par14}, due to the 
larger fraction of stellar emission at shorter wavelengths. 

Variability at different wavelengths (e.g. optical vs IR) and spectroscopic
variability (in emission and absorption lines, affecting the flux or the line profiles) can be very different for the same star. Campaigns to monitor
variability at different wavelengths simultaneously provide
information on the process that causes the variable events (e.g. extinction by disk material, compared to variations in the accretion
rate/accretion luminosity; \citealt{Eiroa00,Eiroa02,moralescalderon09}).

\subsection{Stellar variability and variable accretion}

Accretion variability is a highly debated topic, although part of the discussion 
depends on how one measures accretion: Using the UV-excess as tracer, 
the accretion rate variations {\bf over few-year timescales} are small, typically less than 
0.5 dex both in TTS \citep{sicilia10,costigan12,Venuti15} and HAeBes \citep{Mendi11b,Mendi13}. 
Line variability has been also used to study accretion variations, since accretion-related 
emission lines can change quite
dramatically in flux and profile. However, these lines are affected by other physical 
processes, so even though the variations in line intensity and profile
are stronger in stars with variable accretion \citep{alencar01,herbig08,sicilia12},
rotational modulation of the line intensities and/or profiles is also 
observed in stars with relatively constant accretion \citep{alencar12,costigan14,sicilia12,sicilia15b}. 
This effect has led to proposing stable and unstable 
accretion regimes, depending on whether accretion proceeds via well-defined channels (stable) vs multiple,
quickly changing fingers (unstable; \citealt{kurosawa13,takami16}). 

Without detailed time-resolved data, it is often hard to determine whether accretion itself is variable, or whether there are
rotational modulations due to an irregular distribution
of accretion columns over the stellar surface  \citep{alencar12,costigan12,costigan14}.
Nevertheless, some objects (FUOr and EXOr) display dramatic accretion variations,
with increases in the accretion rate up to a few orders of magnitude.
Both classes were initially identified from optical variability, and 
their nature, evolutionary stage, and the link between them and the rest of PMS stars 
remains open to debate \citep[e.g. review in][among others]{Audard14} 

Besides their interest as part of the star formation pathway and disk evolution, 
stars with variable accretion are key target to understand the physics of disks,
by observing how the disk reacts to a sudden increase in the
accretion rate.
The luminosity variations associated to increased accretion episodes
can dramatically affect the disk properties, processing the amorphous
dust in the disk and creating crystalline silicates \citep{abraham09}, and changing the gas composition in the planet-formation region by evaporating icy bodies and destroying organic molecules \citep{banz12}. An increased disk heating can also shift the water snow line to much larger disk radii during an accretion outburst, as discovered in the FUor disk of V883 Ori \citep{cieza16}, producing ice evaporation and dust fragmentation over a disk region where planets would otherwise typically form \citep{banz15b}.
Subsequent evolution of the disk after outbursts reveals longer-term processes happening in inner and out disk regions, such as the draining of inner disk gas by the central star \citep{banz15a} and the redistribution of crystalline 
silicates to larger radial distances by disk winds \citep{juhasz12}, allowing us to explore disk parameters
that are otherwise hardly observable (disk mass budgets, disk viscosity, radial transport).

\subsection{Time-resolved data as a means to trace disk and stellar properties}

Time-resolved data allows to probe phenomena that involve variability,
but they also allow us to access processes with well-defined timescales,
such as stellar rotation, rotational modulation of accretion,
companion orbits, and disk rotation. It is specially powerful
to track relatively short timescales (hours-days), which typically correspond to spatial scales that
are beyond the current direct resolution limits (e.g. stellar spots, 
accretion columns at few stellar radii, structures and clumps on the inner disk
rim at sub-AU scales).

The puffed-up inner rim may cause a radial part of the inner disk to be in shadow
\citep{Natta01,dullemond01,dullemond04,isella05}. 
Although shadowing by features in the inner disk rim is more common among intermediate-mass
stars, evidence for occultations by clumpy inner disks has been found for objects down 
to the brown dwarf regime \citep{scholz15}.
One common subgroup are the `dippers' or `AA-Tau'-like objects. 
The prototype is AA-Tau, which exhibits periodic extinction events that have been 
traced back to a warp at the interface of the disk and stellar magnetic field that 
periodically obscures they star as it rotates \citep[e.g.][]{bou03}. While this
 behavior is seen more often in the optical (where the extinction is stronger), 
it provides insight on the inner disk since it is the disk that is doing the 
obscuring. 
AA-Tau events occur in $\sim$30\% of the young stellar objects \citep{ale10,cod14}, with rapid 
extinction events associated with obscuration by material in the accretion flow \citep{sta14} 
and longer, quasi-periodic events associated with warps in the disk \citep{mcg15}. 

Quasi-periodic dippers can be used to determine the height of the inner wall, which can
vary by $\sim$10\% from one period to the next \citep{mcg15}. This is similar to 
the `seesaw' behavior seen in pre-transition disks, in which the short-wavelength flux 
increases (decreases) as the long-wavelength flux decreases (increases) \citep{muz09,esp11,fla11,fla12},
which has been successfully modeled by a variable inner disk height.

Azimuthally narrow shadowing structures at ${<}1\,$AU in the optically thick inner disk
height affect the illumination of part or all of the outer disk. They have orbital timescales 
comparable to the light travel time to the outer edge of the disk. As a result, 
the shadows cast by any such structures will be curved, and can take a range of shapes from 
nearly linear to strongly curved spirals \citep{kama16c}. Fitting the shape 
of such features, we can constrain the orbital properties of spatially unresolved structures in 
the inner disk, as well as provide an independent constraint on the vertical structure and 
absolute radial size of the disk.
The $1\,$mag 
variability of the HD~163296 disk in scattered light imaging spanning six years may be direct 
evidence for time-variable shadowing \citep{sitko08,wisniewski08}. 
The systematic study of such variations could constrain the presence and properties of 
large-mass substellar companions and various instabilities \citep{sitko08}. 

Polarization, despite being a main technique to study the extended disk (see Section \ref{scattered-sect}), 
can be also used to study variability of PMS and the innermost disk through unresolved 
photo-polarimetry. Unpolarized stellar light that scatters off dust grains in a circumstellar environment 
will become linearly polarized.  The photometric signal from star and disk combined will be partially 
polarized when the emission from the stellar photosphere deviates from spherical symmetry (e.g. stellar spots), 
when part of the stellar surface is obscured by the disk (e.g. flaring inner disk), or the circumstellar 
disk is inclined or non-axisymmetric, producing polarization levels of 0-2\% \citep[e.g][]{oudmaijer2001}. 
Photo-polarimetric variability can thus reveal changes in the stellar photosphere and the inner disk, 
for example by a warped inner disk which is coupled to the stellar magnetic 
field leading to variable obscuration of the star \citep[e.g.][]{manset2009}.

Time-resolved spectroscopy is a further tool to study the complexity of parts of the star-disk
system that are beyond the possibilities of direct resolution. Applied to
absorption lines, it can be used to track the orbits of atomic gas packages in the
disk and to study atomic gas dynamics \citep{mora02,mora04}.
Bright emission lines
like H$\alpha$ and H$\beta$ can be used to study the stability of accretion \citep{alencar12,kurosawa13}. Metallic emission
lines (Fe I, Fe II, Ti II, Mg I, Ca II, etc.; \citealt{hamann92}) are specially valuable, 
since they are simpler than Hydrogen and Helium emission lines and span
a large range of critical densities and temperatures, which allows to probe different regions within the
accretion structures.
Time-resolved spectroscopy of the metallic emission lines thus provides 
a ``tomographic" 3-dimensional view of accretion, tracing the location, extent,
and physical properties of the accretion columns and associated hot spots \citep{sicilia12,sicilia15b}. 

Time-resolved radial velocity data is also a classical means to determine the presence of planetary and
binary companions. Nevertheless, the signatures of companions in young stars can be mimicked/masked by
the presence of periodic signatures associated to stable accretion columns or stellar spots  \citep{kospal14,sicilia15b}.
This makes it hard, but not impossible (if the star properties are well-constrained), to detect young planets \citep{johnskrull16,donati16,david16}.
Companions embedded in the disk are a potential tool to study disk properties.
Accretion rate 
variations are normally random, but the presence of close-in stellar companions embedded in the disk can cause 
pulsed accretion  (in L54361; \citealt{muzerolle13}) and periodic perturbations to the accretion-related wind  (in GW Ori; \citealt{fang14}).
A companion in the disk acts as a ``disk scanner": It moves through a well-defined orbit that 
crosses (and potentially perturbs) the material in the inner disk. This allows us to use its disturbances to 
estimate the location of the launching point of the disk wind and where the bulk of the matter flow 
runs through the inner disk.

\subsection{Disk dynamics \label{dynamics-sect}}

Radio interferometers not only can trace the spatial distribution of gas within a disk, but also 
its kinematic profile. To zeroth order, the gas in a disk is moving in Keplerian orbits
 around the central star(s). With small corrections due to the pressure gradient and the 
height of gas above the midplane \citep{ros13}, this motion can be used to `weigh' the 
central star(s), as has been done in the AK Sco \citep{cze15} and DQ Tau \citep{cze16} binary systems. 

Beyond Keplerian rotation, turbulence plays a large role in e.g. setting the relative velocity 
of small dust grains \citep{tes14}, regulating the accretion flow through the disk \citep{tur14}, 
setting the vertical chemical structure \citep[e.g.][]{owe14}, setting the minimum mass of gap 
opening planets \citep[e.g.][]{kle12}, among many other effects. In part because of its importance, 
turbulence has received a great deal of attention in theoretical studies 
\citep[see recent reviews by ][]{armitage11,tur14}. The predominant theory for turbulence, 
magneto-rotational instability (MRI) predicts motions of a few tenths of the local sound 
speed in the upper few pressure scale heights of the outer disk \citep{mil00,flo11,sim15}, 
which translates to tens to hundreds of meters per second in the outer disk. 

Given the weakness of this effect and its degeneracy with thermal broadening, observations have been less common than theoretical studies. The high spatial and spectral resolution of radio interferometers can overcome many of these complications, making them a promising source of observational constraints. \citet{sim15} demonstrated, using simulated ALMA CO observations of their shearing-box MRI simulations, that turbulence can change the peak-to-trough ratio of the CO spectrum, as well as the broadening in the images, removing some of the degeneracy with thermal broadening. \citet{gui12} used CS, whose relatively high mass reduces the contribution from thermal broadening, to measure turbulence of $\sim$0.5c$_s$ in DM Tau. \citet{fla15} examined ALMA observations of HD 163296 and put an upper limit of only 0.03c$_s$ on the turbulence in the outer disk, well below theoretical predictions. \citet{hug11} used high spectral resolution SMA observations of CO to derive an upper limit ($<$40m s$^{-1}$) in the TW Hya disk. \citet{teague16} examined ALMA data of TW Hya and tentatively detect turbulence at $\sim$50-130m s$^{-1}$, while emphasising that uncertainties in the absolute calibration are a substantial limiting factor in these types of observations.

Other non-Keplerian effects have also been examined in recent years, often in response to new ALMA observations. \citet{ros14} modeled the rapid radial inflow of gas that can arise in transition disk systems with heavily depleted inner gaps. \citet{cas15} considered a warped disk in which the inner disk is tilted with respect to the outer disk. \citet{dong16} estimates that signatures of gravitational instability should be detectable by ALMA in the gas dynamics of disks up to 400 pc distance. All these situations generate velocity differences on the order of the Keplerian velocity, which can be easily resolved over much of the disk. \citet{sal14} find evidence for molecular winds in CO observations of the binary AS 205 based on the kinematics of the extended emission. Shocks associated with spiral arms can create velocity jumps that are multiples of the local sound speed \citep{zhuetal15} and surface density differences across the spiral arms that are potentially observable \citep{dipierro14,hall16}

The dynamical timescale of the outer disk stretches from decades to thousands of years, while the surface layers can respond rapidly to changes in illumination \citep{chiang97}. This time dependence means that synoptic observations over many years can trace changes in structure as well as the response of the upper disk layers to variable illumination. HH 30, a nearly edge-on disk, exhibits variable reflection from the dust surface layers that can be explained by a central beam of light sweeping through the outer disk \citep{stapel99}. 
Accretion bursts can modify the solid-state features in disks, producing crystalline silicates and
cometary material \citep{abraham09}. These newly-created crystals can be used as a ``contrast" to track
the mass flow within the disk:
Followup of these crystals over months/years can be used to trace transport, mixing, turbulence, and
viscosity through the disk \citep{juhasz12}, even though disentangling the various transport mechanisms
(wind, viscous transport, mixing on the vertical direction, turbulence) is not easy.

As the optical and IR (and soon also longer wavelengths) temporal baseline of observations in public databases grow, 
a new door to exploring long-period variability opens, including studing accretion variability on half a century term,
spectral type changes associated to activity cycles, and constraining the long-timescale variations in protostars and
FUor objects.


\section{Discussion: power and limitations of multi-wavelength observations \label{discussion-sect}}

In this section we discuss what we can learn from combining different observational techniques. We highlight their 
complementarity in terms of disk properties, outline the limitations of individual techniques,
and discuss how the synergy of different observations can help to clarify outstanding problems in our 
understading of disk structure and evolution.

\subsection{Understanding the degeneracies}

Many of the uncertainties in our knowledge of disks result from incomplete 
observations, which lead to degeneracies when interpreting observations e.g. in terms of
disk structure, gas/dust ratio, and dust properties.
When interpreting dust continuum observations such as the SED or the IR visibilities and closure phases,
there is a great degree of freedom in the selection of the dust shape, structure, 
composition  (in particular, for featureless species such as carbon) and dust size distribution used 
to model the data. In addition, gas/dust ratios are usually unknown.
Consequently, models are intrinsically degenerated: a large number of models may
describe the data equally well. An additional complication comes from the parameter space 
being strongly non-continuous, so very different geometries, dust properties, 
and structures may provide a similarly good fit.

The stellar properties have a large impact on the disk structure as well as on its observables. 
The strong UV and X-ray irradiation by the young central
star creates complicated and quite unique non-LTE conditions that cannot
be studied elsewhere, which is one of the difficulties in
understanding disk chemistry. In addition, a good characterization of the star (spectral type, extinction, luminosity, activity)
is the only way to observationally constrain 
many disk properties, such as ages and accretion rates \citep{sicilia06b,sicilia10,manara12,manara13,dario14}.
The observational impact is larger for the
parts of the disk SED for which the star provides a strong contribution, such as the NIR 
and the UV. The intrinsic variability of young stars is another potential source of inconsistency between non-simultaneous datasets.

Moreover, since disk emission is essentially stellar light
reprocessed by the disk material, the central star plays an important role on observations, including the 
sensitivity (disks around late-type stars are significantly fainter at short wavelengths than 
their HAeBe counterparts, even if the disk masses are similar), contrast 
(e.g. detecting UV excesses and line emission from accretion in intermediate-mass stars is 
harder due to the higher continuum and deeper photospheric lines), and technical feasibility (e.g. NIR 
interferometry and scattered light observations require a certain
magnitude for the central star). This produces observational biases with respect to the kind of disks we observe for stars with different
masses, regarding for instance the detectability of holes and gaps, which need to be taken into account before deriving general
properties and timescales.

The dust grain properties are a source of uncertainty and degeneracy for both unresolved and resolved
observations. Usually, dust grains are assumed to have a distribution of sizes between a maximum and a minimum size, with a 
power-law exponent with slope -3.5 (from collisional equillibrium). This relation can nevertheless change in 
protoplanetary disks: grain sticking and fragmentation, together with the special physical conditions (very different from
the ISM), can affect the grain properties and emissivity \citep{brauer08,zsom11}. Dust size distributions can furthermore 
vary radially in the disk and with respect to the midplane due to size-dependent filtering by planets/companions
and settling.
The degeneracy on the dust properties can be reduced by modeling additional data sets.
For example, a 10$\mu$m spectrum covering the silicate emission can constrain the dust composition and size distribution in the upper layers of the disk.
Scattered light images can constrain the amount of dust settling \citep[e.g.][]{Pinte08,pinte09,Pinte2016}, and
sub-mm images constrain the mass of dust in large grains in spatial scales similar to the beam size,
as well as the grain size and disk mass.

Understanding the gap physics 
from observations of individual disks with known planets could be
also the starting point to interpret the observations of gaps and holes in other disks where the presence of planets is
not known.
Sparse aperture masking can be used for the detection
and imaging of young planets and circumplanetary disks, and for the detection of complex structures within the disks (Biller et al. 2014).
Angular differential imaging (ADI) has also been used to search for protoplanets \citep{quanz2013}.
When accretion-related lines like H$\alpha$ are observed with ADI, we can also detect accretion
onto the protoplanets and study the structure of the gas within the cavity \citep{sallum15}.
Spectro-astrometry of CO lines can be also applied to the search for planets and associated disks (\citealt{brittain15}; but see
caveats in \citealt{fedele15}).
Spectro-astrometric signatures of accreting protoplanets should be detectable in optical and NIR emission lines 
(H$\alpha$, Br$\gamma$). NIR data, where the contrast with the star is lower, is very promising,
athough X-shooter observations have not been successful so far \citep{Whelan15}.
Similarly, detailed studies of accretion and activity can help to improve  our knowledge of these processes, helping to distinguish
their effects on radial-velocities and photometry from those of planetary companions \citep{sicilia15b}.

We still need to be careful with the difference between what has been determined about 
disks, and what relies on inferences. Some uncertainties can be addressed by an improvement in
observational capability or in models. For example, with very high spatial resolution, 
we can determine a scale height directly by observing an edge-on disk in a tracer of 
choice. Or, if we need to know if a molecule varies greatly in abundance within a disk, 
we can apply a suitably large chemical network. However, other problems remain where we 
need to make sure that our inferences do not become established `facts'. Examples 
include the masses of disks and the signatures of planets, where our actual knowledge 
is much less sophisticated than, for example, the directly measurable properties of 
stars and exoplanets.

\subsection{The power and limitations of unresolved, multiwavelength data}

\begin{figure*}
\begin{center}
\includegraphics[width=15cm]{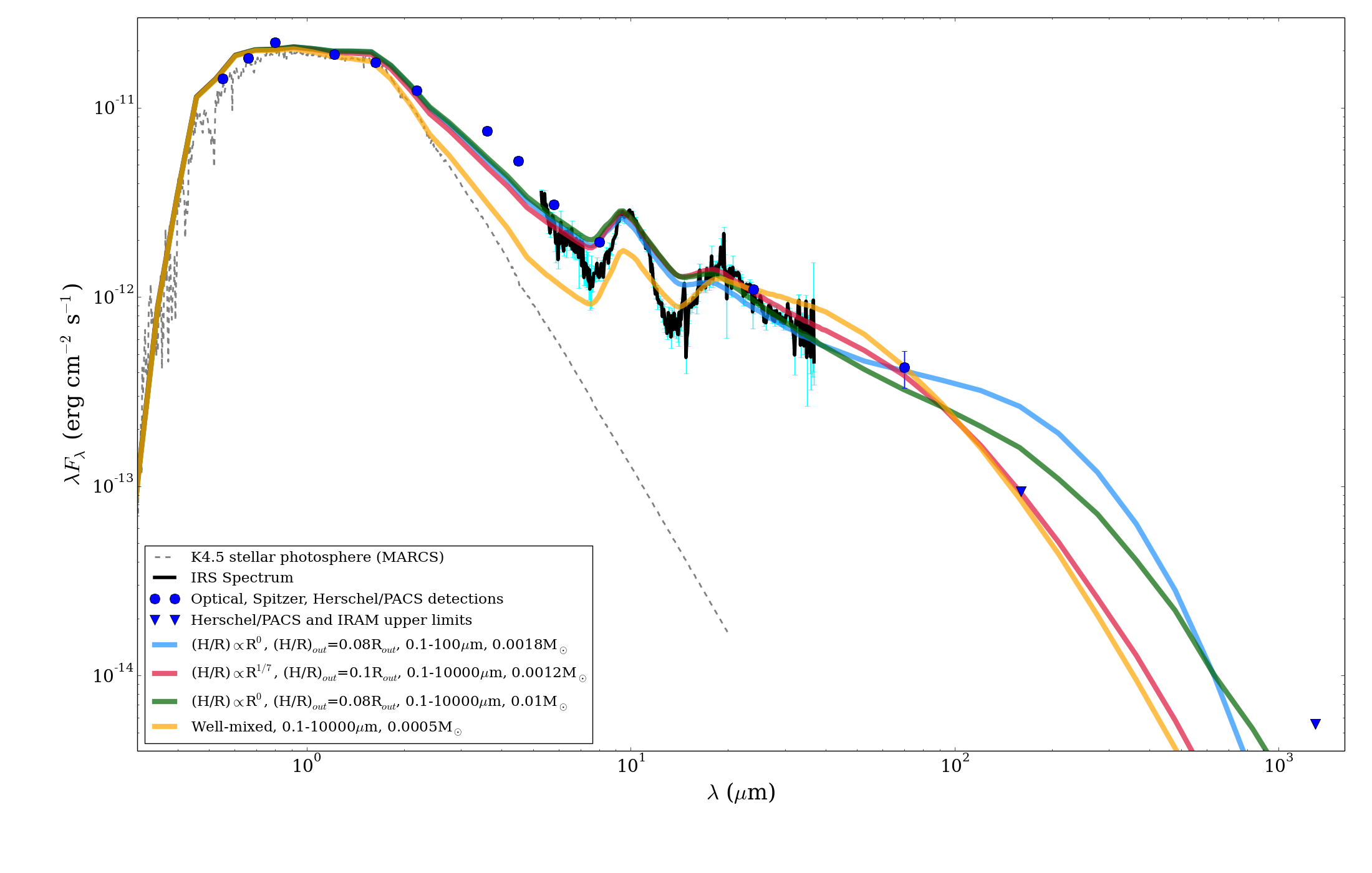}
\caption{An example of how a very different (and non-continuous) distribution of disk models, including changes in disk mass, 
vertical structure, grain sizes, and settling) can provide 
similarly good fits to partial multiwavelength data \citep[based on][]{sicilia15a}. The more data we include, 
the more models we can rule out. For the present example, NIR data can exclude a well-mixed gas and dust disk model, while
the far-IR data and mm-wavelength upper limits put a strong constraint to the dust content in the disk. The fact that the
silicate feature is not well-reproduced by any radially continuous disk model further indicates the presence of unresolved holes or gaps,
which were not included in the model. }
 \label{sedexample-fig}
\end{center}
\end{figure*}

One classical example of analysis of observations that is affected by multiple degeneracies is SED modeling and interpretation.
The first problem is the lack of information on the gas, beyond what can be indirectly deduced from indicators such as accretion or disk flaring.
Even though a good characterization of the star
and a relatively complete multiwavelength dataset can exclude many possible disk structures, well-fitting
detailed SED models for individual objects are often later contradicted by spatially-resolved observations. 

The broadband continuum SED of a disk can constrain the total dust mass up to 
particle sizes comparable to the longest observed wavelength, as well as the degenerate combination of gas 
disk flaring and large-grain settling.
Although determining the size of an unresolved hole in the inner disk is strongly dependent on the assumed
dust properties (including grain size) and shape of the inner rim \citep{calvet02,ratzka07,sipos09,sicilia11}, a complete optical and
NIR SED can prove that a radial variation on disk properties (mass, grain sizes, and/or vertical scale heights)
is needed to explain low IR fluxes \citep{furlan05,dalessio06,sicilia13a}. Similarly, even though the disk mass and scale height are highly 
degenerated if no millimeter data is available, there is a strong correlation between MIR data and millimeter observations 
\citep{lommen10,currie11}.

Far-IR data can provide strong constraints to the disk structure, specially important in case of faint 
and evolved disks \citep{sicilia15a}. 
The shape of the MIR SED depends on both the disk flaring angle and the presence of cavities. 
MIR colors are in good correlation with the amount of detected scattered light, suggesting that both
depend on the disk scale height. However, it is not known how often these disks may have non-detected 
gaps, since most (all?) of the resolved disks have gaps (Section \ref{gaps-sect}).

SED interpretation relies on disk models, which can be applied to dust and gas observations. Codes such as, but not limited to, 
RADMC \citep{dullemond04},
DALI\footnote{\texttt{http://www.mpe.mpg.de/$\sim$simonbr/research\_dali/index.html}} \citep{bruderer12,bruderer13}, 
PRoDiMo\footnote{\texttt{http://homepage.univie.ac.at/peter.woitke/ProDiMo.html}} 
\citep{woitke09}, MCFOST \citep{pinte06}, and MCMax \citep{min09} allow to model disks and to reconstruct observables (SEDs, resolved images)
that can be directly compared to the data. Some models can simultaneously fit the SED, spatially resolved data, and atomic and molecular line fluxes. 
Published grids of models give guidance on how to combine the 
SED, line fluxes and profiles, and spatially resolved data to constrain various structure parameters, 
gas masses, snowline locations, and elemental abundances in the inner and outer disk 
\citep{gorti04,robitaille06, woitke10,kamp11,qi11,woitke16,kama16a}
A number of studies have fitted multi-wavelength data to determine the disk structure, gas mass, and C and O abundances 
\citep{gorti11,tilling12,carmona14,du15,kama16b}. 
Care has to be taken when fitting disks with a pre-existing model grid, as this will not be
able to pick up structures or disk properties that had not been previously included in the grid.
Statistically similarly ``good" fits may miss key features (e.g., the silicate feature in the SED example in Figure \ref{sedexample-fig}),
so ruling out disk structures incompatible with the observations is often a more powerful way to extract information from multiwavelength
SEDs than finding a good (but not unique) match \citep{sicilia11,sicilia15a}.

In summary, combined with statistically significant observations of
many stars, SED modeling and interpretation is one of the most time-inexpensive methods to characterize
the general properties and evolution of disks, especially faint ones around solar-type and low-mass objects.
When large samples are included, the average global
properties and timescales deduced provide a representation of the general behavior.
Even if the interpretation of individual objects may remain ambiguous, 
a fine-detail model is not always the answer: for instance, if we want to obtain a global view of the role of planets in disk dispersal,
it may be enough to know what percentage of
disks have gaps or holes consistent with planet formation, even if we do not know the precise sizes and
locations of all the gaps and planets.

\subsection{Planets and asymmetries: Intrinsic structure or evolution?}

While there is a general consensus that the typical lifetime of disks is a few Myr, the way disk disperse and evolve and
the observational signatures of disk dispersal processes are still uncertain. Recent observations show that it
may be very hard to draw a clear line between disk structure and disk evolution. 
Disk gaps and holes are ubiquitously found by the increasing number of spatially- or spectrally-resolved observations, 
so that ``classical", primordial, continuous disks may indeed be the exception.
This is especially true for HAeBe stars, for which
both continuum and scattered light images show that almost all disks are asymmetric (Sections \ref{innerdisk-sect} and \ref{gaps-sect}),
but the (few) available high-resolution data on low-mass stars reveal potential evolutionary signatures, rings and gaps, 
as well (e.g. HL Tau, TW Hya).

Radial asymmetries may form earlier than previously thought in the disk lifetime, even if 
most unresolved observations such as broadband SEDs are compatible with radially continuous disks. In this case, 
the ``transitional" phase of disks would be harder to define, as most disks would have holes/gaps/asymmetries from early on.
The rate of structural changes in disks is essentially unknown, except in some extreme and rapidly variable cases  
\citep[e.g. the triple system GW Ori, whose inner hole has suffered several dramatic changes in the last $\sim$40 years;][]{fang14}.
The high frequency of gaps and holes observed would thus indicate that they are either produced all the time, or that they can survive
for a few Myr. 

Detailed, spatially resolved data and multiwavelength observations of large samples of young disks around stars 
with different masses, will be a key to investigate which gaps and holes
imply evolution (and which objects are the best candidates to be ``about-to-disperse" or ``in transition"), 
and which may result from long-lasting processes operating through most of the disk lifetime.
A comprehensive picture is still lacking but an effort to understand whether disks with(out) cavities, 
spirals or rings, extended vs compact structures, are different evolutionary stages or evolutionary paths starts to be 
possible \citep[e.g.][]{maask13,sicilia13b}. Surveys that include disks around 
low-mass stars and around intermediate-mass stars will
help to distinguish differences and similarities in their evolutionary paths \citep{BP15}.
Being able to detect and resolve evolved disks around low-mass
stars will also help to check whether gaps in disks around low-mass and intermediate-mass stars are similar entities,
created by the same physical processes.

The morphology (and possibly also the nature) of the asymmetries observed in continuum and scattered light are different. The different
coupling of  
small and large grains with the gas 
\citep[which includes the possibility that large grains are not coupled at all;][]{laibe12a,laibe12b,dipierro16}
 is the likely cause, and can be used to explore the
causes behind the asymmetries. Searching for counterparts of 
the scattered-light spirals and asymmetries at mm wavelengths can reveal whether these structures are ripples
in the height distribution of micron size grains, or if they affect the whole disk, including larger grains \citep{akiyama16}.
Asymmetries affecting all disk components (gas, dust with different sizes) could be a signpost of processes 
such as gravitational instability \citep{durisen07}, which involves the entire disk structure. 

\citet{Pinilla2012b} and \citet{dejuanovelar13} showed how the discrepancy between the 
location of the disk inner edge for micron- and mm-sized grains can be ascribed to the dust 
differentiation from the interaction with planets. The first observational proofs were obtained by
\citet{garufi2013}, later confirmed by others. \citet{dong12} showed that in some 
cases, a hole/gap for micron size grains may not even exist at all where seen for mm grains. 
Thus, a clear link between observed brightness decrease and intrinsic disk mass depletion has not yet been established,
at least in scattered light. 

Many authors have studied the effects of planet-disk interaction in the context of spirals
\citep[e.g.][]{juhasz15,pohl15,Dong2015,Zhu15}, although often the models do not provide fits with 
plausible physical conditions and/or perturber mass. No planets have been detected in spiral disks yet. More in general, 
there is no disk feature that could unanimously be ascribed to the interaction with known planets, even 
though the effects of known companions in disks holes and gaps has been documented for a few objects \citep{biller12,fang14}.
Moreover, most of the transition disks studied in scattered light may be outliers in the 
nominal evolution of disks (and we are thus biased), otherwise it is hard to reconcile the observed 
planetary system architectures with the disk observations \citep[e.g.][]{dong16}. 
Finally, the aspect ratio and the location of the disk inner edge and its brightness profile can reveal 
information on the cavity nature, using scattered light \citep{quanz2013,garufi2014, avenhaus14b}
or MIR interferometry \citep{mulders13}. 

In addition, there is a wide range in planetary mass and size that we cannot directly detect. The 
largest radio-detectable dust grains are centimeters across, while an accreting 
proto-planet will have a Hill sphere on AU scales. The only detectable sign of 
intermediate-mass planetesimals is perhaps in `falling evaporating bodies', i.e. 
exo-comets that leave a redshifted spectroscopic signature towards the host star. 
In late-stage `debris' disks, planetesimals should be present to regenerate dust that 
is short-lived compared to the host star; however, backtracking to the planetesimal 
sizes and locations is difficult. Further, some key physics on e.g. gas viscosity, 
turbulence motion and ice formation is unclear, so that infering planet presence from 
apparently cleared gaps and cavities in disks is risky. In the future, observing forming planets 
-- perhaps with JWST infrared imaging --, or being able to apply usual
planet detection techniques to young stars (which requires a very good understanding of the
effects of accretion and activity on the planet tracers) may allow us to determine what 
actually makes a disk `proto-planetary'.

\subsection{Disk mass and disk evolution \label{diskmassdiskevo}}

Although timescales favor the dispersal of protoplanetary disks from the inside-out \citep{hayashi85,strom89},
some disk dispersal processes such as gravitational instability may be more efficient in the 
outer, colder parts of the disk \citep{boss97,rice06}. Moreover, observations of globally mass-depleted disks
\citep[with low small-dust content;][]{currie09}, extensive grain growth  \citep{rodmann06,ricci10},
significantly settled disks \citep{furlan05,dalessio06,sicilia11}, and gaps at large distances \citep{ALMA2015} suggest that
a large degree of evolution may have happened through the whole disk before it starts to dissipate from the inside-out.
A variety of 
evolutionary paths could also be behind the diversity of exoplanetary systems.
From statistical studies using multiwavelength, unresolved data, 
TTS appear to follow several distinct evolutionary paths.
Transition disks with inner holes appear in two flavors: accreting and non-accreting
\citep{fang09,sicilia10}. Besides the presence/lack of accretion, the two classes also differ in the
small-dust content of the disk \citep{sicilia13b,sicilia15a}. 

The accretion behavior (including lack of
accretion) among small-dust depleted disks is substantially different from that of primordial
and non-depleted disks \citep{sicilia13b}. In fact, non-accreting disks are
exceedingly rare among primordial and transition disks that are bright in the far-IR \citep[and thus
massive and/or flared and gas rich;][]{sicilia15a}.
Therefore, having a hole is usually not
sufficient to shut down accretion, while having a low small-dust mass seems to affect accretion independently
of the presence of (SED-inferred) holes. A low disk mass is also not a requisite to open a hole.
This diversity among dispersing disks suggests that the opening of inner holes may occur at different
evolutionary stages \citep{sicilia13b,sicilia15a}, as it would be expected from the interplay of different disk dispersal 
mechanisms (photoevaporation, planet formation, viscous evolution). 

Among TTS disks with holes inferred from their SEDs, 
accreting and non-accreting transition disks have by definition similar NIR and MIR colors
and excesses, but their Herschel far-IR fluxes are surprisingly and significantly
different. Compared to primordial disks around stars with similar spectral types, 
non-accreting transition disks have clearly lower 70$\mu$m flux, and accreting transition
disks have higher 70$\mu$m fluxes\citep{sicilia15a}.
These differences in far-IR colors
suggest that non-accreting disks have significantly
lower masses and/or clearly flatter disks \citep[which would also imply small-dust depletion, extreme settling and probably, gas depletion;][]{sicilia15a}.
Low disk masses in non-accreting disks are in agreement with the predictions of photoevaporation
as a mean to stop accretion once the disk mass (and thus the viscous flow through the disk) have 
decreased enough \citep{gorti09, sicilia10}.
The higher 70$\mu$m fluxes of accreting transition disks (compared to primordial ones) could indicate that inner
holes are more likely formed in disks with a high mass, although other effects such as changes in the
vertical scale height (for instance, induced by further gaps at larger distances) could also contribute to
higher 70$\mu$m emission \citep{sicilia15a}.
These differences in far-IR fluxes among TTS disks may be the analog to the mm-bright and mm-faint disks \citep[or group I/II HAeBe disks;][]{meeus01} 
among HAeBe.

The total disk mass would thus be an important parameter in
evolution, in addition to (and independently of) the presence of inner holes or gaps. 
Given the abundance of disks with asymmetries as shown by resolved observations, mass depletion
may be a better signature of imminent disk dispersal than holes and gaps. 
The fact that accretion strongly decreases (or even ceases) in low-mass and 
dust-depleted disks is a sign that the transport through the disk
fails as the disk loses mass, eventually leading to the rapid dispersal of disks.

In this case, the uncertainty in mass estimates is a strong limitation to understand disk dispersal.
Figure \ref{massunc-fig} has illustrated orders of magnitude difference in 
derived mass from different methods and assumptions. We are forced into this situation 
because the bulk of the mass is in molecular hydrogen, and only 4 disks have been observed in
directly-related HD (Section \ref{gasmass-sect}). There are very good reasons to think 
that the gas-to-dust ratio and abundances of trace gases vary, between systems and 
across individual disks. So for the ensemble of disks, some other `reality check' 
arguments are important. Observations that can indirectly probe the global disk mass and gas content, by detecting processes that connect or involve the whole disk, 
such as accretion, or by studying the coupling between gas and dust (e.g. studying the disk vertical scale height for different grains) can 
help to pin-down the disk mass. Further sanity checks on our understanding of disk masses include checking whether the observed mass budgets sufficiently high to form 
the ensemble of observed exo-planets, to feed accreting protostars (specially important if planets form very early on, as suggested by HL Tau), or low enough 
that the star-disk systems are gravitationally stable. Applying such logic may help to
understand why some tracers are better than others, as well as to devise a better way to estimate the total disk mass.
The high sensitivity of modern interferometry will be a key to detect faint, low-mass disks (usually too faint for single-dish mm-wavelength observations) 
in both gas and dust. If a low disk mass is found to be the main key to a rapid disk dispersal, observations of
low-mass disks will be a challenge for the future, including future instrumentation.

\subsection{Gas as a probe of disk evolution \label{gasdustprobes-sect}} 

In the disk environment, the physics of the gas and dust strongly depends on
the interactions between the two components. While many problems arise when trying to connect gas and dust (differences in disk masses and sizes estimated from gas and
dust, dust decoupling and settling, different behavior of gas and dust in planet-related gaps, etc), the gas/dust
connection is a useful tool to trace processes that are otherwise hard to observe.
In particular, the gas density distribution inside the dust gaps can provide information on their origin:
The two leading mechanisms for gap formation, photoevaporation \citep[e.g.][]{clarke01, alexander06, ercolano09,gorti09}, and dynamic interaction 
with giant planets \citep[e.g.,][]{rice06}, predict clear differences in the evolution of the gas surface density in a disk.

A Jupiter-mass planet would quickly open a gap in the large dust grains at the location of the planet (about $<$2-4 AU width, depending on the planet mass),
but blocking the gas and the well-coupled small dust is harder.
As the disk evolves, the disk's surface density at radii smaller than the planet would decrease with time, eventually creating a gas density drop
\citep[see for example Figure 5. in][]{tatulli11}. Large planets can produce gaps in gas and dust, but smaller ones only create dust gaps and, in general,
dust gaps are highly planet-size dependent \citep{dipierro16}.
Photoevaporation would quickly open a several-AU gap in the gas at the location of the critical radius.
The gap would grow in a very short timescale to sizes $>$5 AU, while the inner most disk 
would lose gas due to accretion and photoevaporation, until accretion would terminate once the innerdisk  gas has been accreted. 
Nevertheless, the two mechanisms are likely to happen simultaneously.
If combined with
giant planet formation, X-ray photoevaporation may be responsible for the origin of transitional disks with large inner holes
\citep{rosotti13}. Photoevaporation
may also explain why stars with very low accretion rates among solar-type objects are rare \citep{sicilia10}. 

The observed dichotomy for transition disks around low- and intermediate-mass stars offers an
interesting comparison with this bimodal way of disk dispersal.
\citet{Owen2016} discusses the possibility of having two families of holes produced by different mechanisms (planet vs photo evaporation)
among the mm-bright and the mm-faint HAeBe disks. Possibly all the dust gaps and dust holes seen in transition disks 
around HAeBe stars have gas inside  \citep[e.g.][]{pont08,BP15,vdplas15}, especially for dust gaps resolved at sub-mm wavelengths. Current observations of gas inside the cavities of transition disks \citep{carmona14,carmona16,Bruderer2014,BP15,vanderMarel2015-12co,vanderMarel2016-isot} 
provide evidence for gas density drops in transition disks.
Nevertheless, all gaps seen in intermediate-mass mm-bright transition disks are very large ($>$10 AU) and 
significantly different from the small dust cavities found in mm-faint disks. 
The mm-faint disks of HAeBe
would be the higher-mass analog of mass-depleted or anemic transitional disks \citep{lada06,currie09}, which also appear significantly different in terms of accretion and
gas content \citep{fang09,sicilia13b,sicilia15a}.
The bright, accreting transition disks around TTS would be the equivalent to the mm-bright HAeBe transition disks.
The parallelism is further extended considering that, as mentioned in Section \ref{diskmassdiskevo}, accreting and non-accreting transition disks show different 70$\mu$m emission, 
by amounts that require a strong dust depletion and not only a change in vertical structure \citep{sicilia15a}. 
For HAeBe, the gas properties of transition and primordial disks are different. 
The fact that accreting TD have lower accretion rates than primordial disks with the same mass may be pointing in the same direction \citep{najita07,sicilia15a,najita15}.

Connecting the disks around TTS and HAeBe would require unifying their observations. The main problem to trace the gas throughout the disk 
is the dependency of gas tracers on the temperature profile (and of the temperatures on the stellar mass, as shown in Figure \ref{innerrim-fig}) and the solid angle.
The correlation between CO ro-vibrational emission and the CO (sub-)mm emission in primordial and transitional disks \citep{woitke16,carmona14}
suggests a connection between the inner disk gas and the outer disk gas. Although millimeter CO ALMA data can detect the faint emission of colder gas down to 5--10 AU, sensitivity is a problem for high-resolution observations, so observations of the ro-vibrational CO transitions are much more sensitive to gas at 0.05-20 AU \citep{BP15}.
 The combination of the two techniques can be very instructive, providing a unified view of gaps from the smallest to the largest. 
By comparing the inner radii of CO ro-vibrational and CO rotational we can also distinguish gaps from holes with no gas inside the dust gap.
Including accretion and atomic gas observations, the presence of gas can be thus traced from the outer to the innermost disk even if the CO ro-vibrational emission is very weak.

In summary, gas observations favor planet formation as the dominant scenario for the formation of the dust cavities and gas density drops in bright and massive transition disks.
However, the large density drops observed in the gas around HAeBe, and the lack of accretion in a significant number of transition and dust-depleted disks around
low-mass stars, are more easily explained by a combination of planet and photoevaporation than by the presence of a giant planet alone \citep[see][]{owen15}.

\subsection{The time dimension}

\begin{table*}
\caption{Key observations and the processes that they can help to distinguish. The sections where the corresponding discussion can be found are also listed. }              
\label{obs-table}     
\centering                                     
\begin{tabular}{l l l}       
\hline\hline                        
Observations &  Disk parameter/Physical process & Sections\\ 
\hline
Accretion and hole properties & Photoevaporation vs planet formation & \ref{accretion-sect}\\
Accretion and disk mass &  Matter transport, viscosity & \ref{accretion-sect}, \ref{diskmass-sect}\\
Radial size of gas and dust disk & Grain growth, viscous evolution & \ref{innerdisk-sect},\ref{gaps-sect}\\
Presence of gas/dust in holes & Matter transport, viscosity & \ref{gaps-sect},\ref{accretion-sect}\\
Spatially-resolved grain sizes & Grain growth, dust trapping & \ref{gaps-sect}\\
Time-resolved stable/unstable accretion & Transport in disks, stellar magnetic field & \ref{accretion-sect},\ref{timeresolved-sect}\\
Scattered light and mm continuum & Disk scale height, gas/dust mixing & \ref{gaps-sect}\\
Mid- and far-IR/mm data on holes & Mass in the holes, photoevaporation, planets & \ref{gaps-sect},\ref{accretion-sect}\\
Time-variable shadowing and scattering & Global disk structure and vertical scale height & \ref{gaps-sect},\ref{timeresolved-sect}\\
Low-metallicity disks & Gas/dust connection and observational relations, viscosity & \ref{diskmass-sect}\\
\hline                                             
\end{tabular}
\end{table*}

As the usual dichotomy between single-object detailed observations and multi-object unresolved datasets
is disappearing thanks to the gain in sensitivity of current instrumentation, the potential of using 
``special" systems to track certain processes is also gaining acceptance. 
``Special objects" (with variable accretion, with anomalous disk masses, with anomalous metallicity, with extreme inclinations) 
can be used as ``stellar experiments" to explore
the parameter space in unusual conditions (e.g. how the disk and the star react to a sudden increase in the accretion rate, the differences and origins of
disks with very different masses around similar stars, the role of dust and its coupling to gas in low-metallicity disks).

Strongly connected to the previous point, the time-dimension is another of the most promising lines to explore in the coming years. 
To every timescale there is a spatial scale, which can be smaller than what can be currently resolved by other means.
Besides obtaining new time-resolved data, we are reaching a time when good-quality observations are available
for many objects, covering the past $>$40 years. The availability of high-quality optical and NIR photometric data since the early 70's has already
allowed us to explore timescales in multiple stars with disks comparable to several orbital periods, unveiling how the companion-disk interaction
clears the disk and filters the dust \citep[e.g. GW Ori;][]{fang14}. 

As high-resolution observations with ALMA accumulate year after year, it will become possible to study disk dynamics in the planet-forming
regions (with 10-20 years periods).
Periodic phenomena can help to understand disks and trace parts of them that 
cannot be resolved by other means. For instance, using known companions as \textit{scanners} that move through the disk \citep{fang14},
or using observations of stable/unstable accretion. The dynamics of the accretion columns can be used to indirectly explore the stellar magnetic field
configurations, in stars where these direct observations are hard due to accretion and activity \citep{sicilia15b}.
Periodic or quasiperiodic disk obscurations (including dippers) are also a key to explore the dust properties and the disk scale-height, which can
help us to understand and interpret other systems even if they do not have extreme inclinations.

\subsection{A protoplanetary disk ``Rosetta Stone''}

From the discussions presented in previous Sections, the power of combined 
multiwavelength observations emerges as a way to reduce the degeneracies in the interpretation
of disks in terms of disk physics and structures. This power to unravel
disks is based on the different ways in which disk physics affects the various disk components, such as gas and dust, which are interconnected by being
parts of the same structure: the disk.  Wherever results from different wavelengths appear incompatible or contradictory, 
rather than a problem or conflict, it
should be considered as an opportunity to explore new disk physics. 

The complementarity of the various observations discussed in the previous sections is
shown in Figure \ref{rosetta-fig}, which we call \u201cthe Disk's Rosetta Stone\u201d as an attempt to identify 
common physical processes that can be decrypted and investigated through different wavelengths.
Combined high-sensitivity observations allow us to access some parts of the disk
parameter space that are not directly observable (such as the disk viscosity or turbulence). 
In Table \ref{obs-table} we offer a summary of some of the key complementary observations
arised in our analysis and lists the relevant sections. More details are given below:
\begin{enumerate}
\item Observing accretion and/or gas within dust inner holes can help distinguishing inner holes related to photoevaporation vs those opened by other processes
(e.g. planets). Combined with disk mass observations, we can extract information on the disk viscosity and also detect other processes that may interfere with
gas transport in the disk (e.g. planets/companions, gravitational instability).
\item Exploring the sizes of disks in gas and dust can reveal to which extent viscous evolution of the disk has altered the initial well-mixed state. 
If data on grains with different sizes is included, grain growth and size-dependent dust filtering can be also probed, including their radial dependencies.
\item The differences in the gas and dust content within gaps are also related to disk viscosity (a parameter that governs the transport of matter through the disk) and to the coupling between gas and dust. Direct resolution of gas and dust in disks with different masses or at different radii can help to map the viscosity through disks.
\item Time-resolved observations of accretion tracers (especially with spectrally-resolved emission lines, although photometric studies of hot spots could also be used) can reveal the stability of accretion as well as the physical properties and distribution of accretion columns. Given the connection between accretion and the stellar magnetosphere, they can be
used as indirect probes of stellar evolution.
\item Combined scattered-light images and dust continuum SEDs can constrain the disk vertical scale height and gas content. Large grains can be detected
with sub-mm data.  From the differences between large and small grains
and gas, we can also gather information on the gas content, disk flaring and dust settling. Silicate feature observations can complete the surface dust picture.
\item Mid- and far-IR data can be combined to distinguish between mass-rich/gas-rich/flared and mass-depleted transition disks that have essentially the same NIR colors.
Exploring the structures of these disks can unveil their evolutionary stages and the processes responsible for their radial asymmetries.
\item Observations of regions with various metallicities can be used to explore the gas and dust interconnection, as well as our relations and calibrations between
different gas and dust components.
\item Time-variable shadowing by the disk, which can be observed via scattered light imaging or resolved mm observations, 
can be used to explore the disk structure and scale height. The
periodicity (or quasi-periodicity) of the obscuring phenomenae and the light time travel to track distances across the disk
are keys to determine the radial structure of the disk.
\end{enumerate}

\begin{figure*}
\begin{center}
\includegraphics[width=17cm]{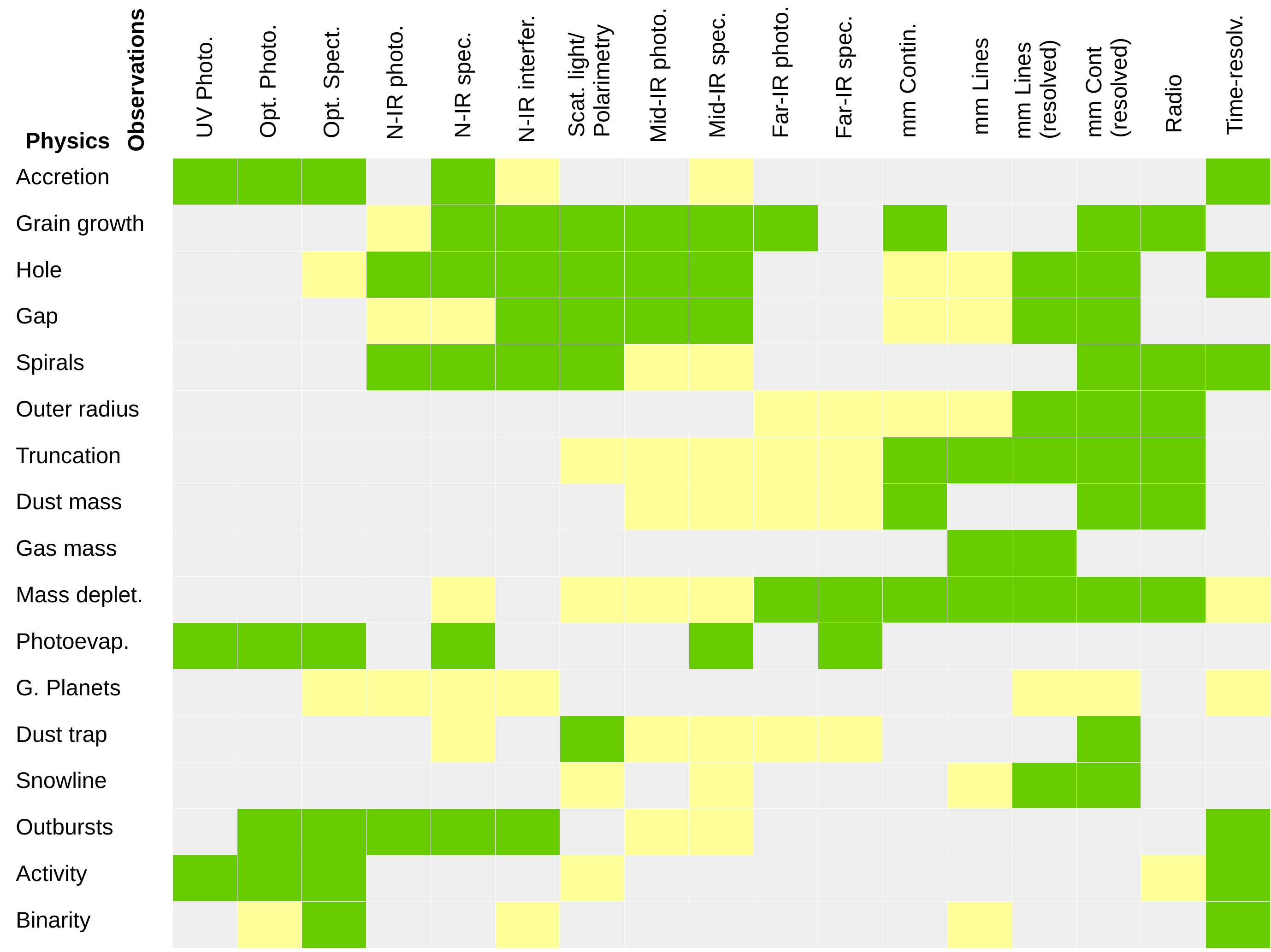}
\caption{The disk Rosetta Stone. The top row lists the available observations, the left column lists 
different aspects of the disk structure and evolution. Green cells mark the places where observations
can provide clear information about the given aspect of the disk structure. Yellow cells mark observations that provide some
information on the disk structure, but subject to different interpretations in terms of disk models and/or disk
physics. The combination of multiple observations allows us to trace the various parts of the gas and dust 
in the disk, connecting them through the different physical processes happening in disks.}
 \label{rosetta-fig}
\end{center}
\end{figure*}

One of the main limitations at present is that the coverage of multiwavelength data is very limited: there are very few
objects with complete and spatially-resolved datasets, and they tend to be massive, bright disks which may not be representative for the origin of 
most planetary systems (including our own). New instrumentation, with higher resolution and higher sensitivity, is currently closing-in the gap between detailed
studies of single and bright objects, vs less detailed observations of statistically significant samples of disks, as the summary table in Figure \ref{feasibility-fig}
shows.

\begin{figure*}
\begin{center}
\includegraphics[width=17cm]{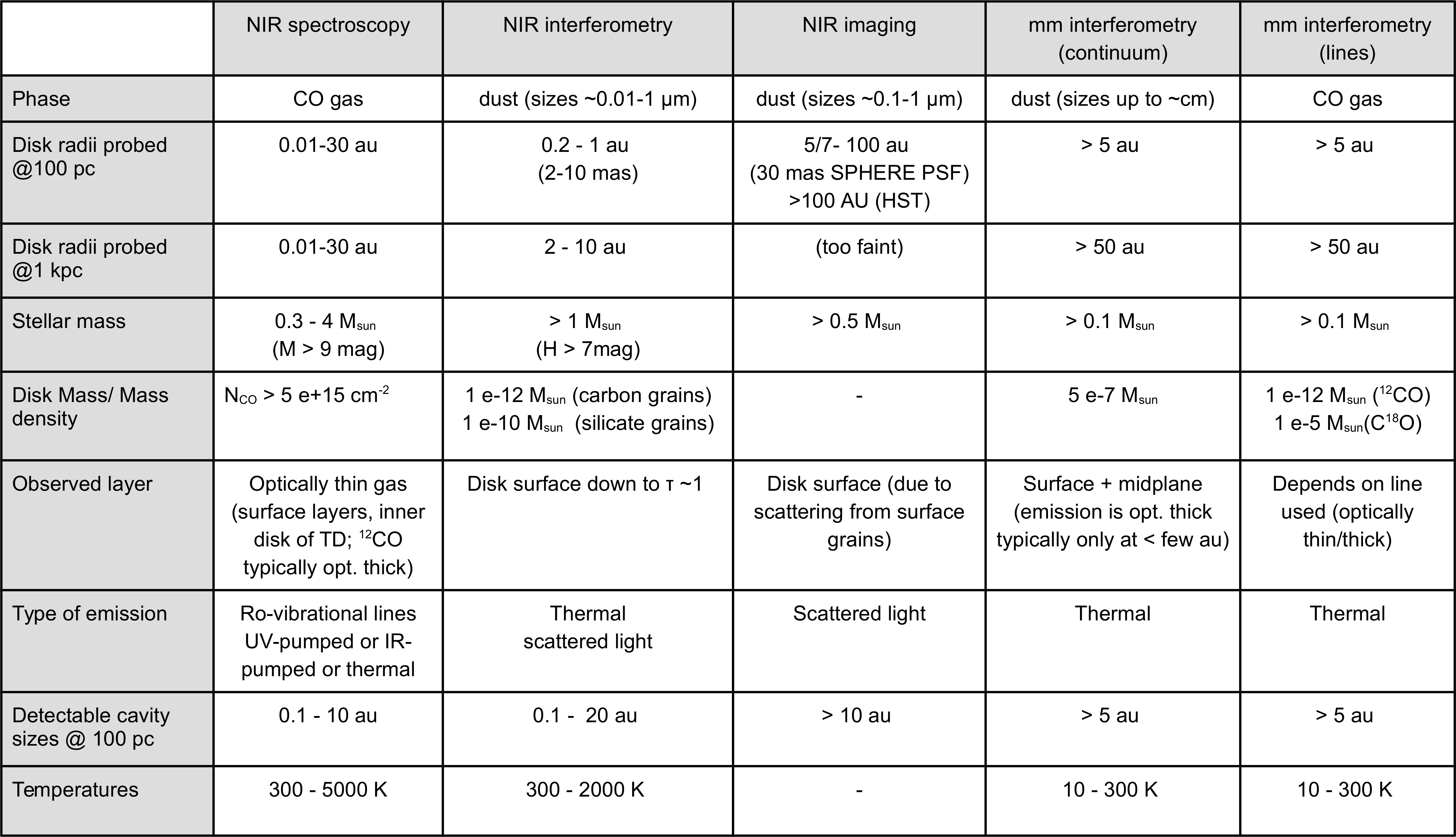}
\caption{Feasible observations with current instrumentation. The table summarizes the kind of objects and disk regions that can be
currently probed with the available instrumentation, depending on their distances, stellar, and disk properties.}
 \label{feasibility-fig}
\end{center}
\end{figure*}

As very detailed observations of many disks start being a reality, finding common points between disks and connecting observations
either as results from the same physical process (with different initial conditions) or as different time-steps on initially similar
objects will be also part of the analysis and interpretation of disks. From what we have learned from unresolved observations, 
the evolution/structure connection may not be evident nor immediate, so observing statistically significant samples will become a necessity.
Future 
telescopes will also help in this direction by improving efficiency and sensitivity, allowing us to connect what we can observe in bright, nearby objects and
fainter ones (including low-mass and solar-type stars, as well as more evolved disks). This will help to understand disks
in their full range of stellar mass and time evolution.

Similarly, better models that allow for 3D disk structures will become more and more needed due to the increasing availability of
spatially-resolved observations. Although the present paper does not address the complex disk chemistry, high-sensitivity and spatial resolution of
molecular line observations will be an additional powerful tool to explore disks (see \citealt{haworth16} on challenges for disks modelers).


\section{Conclusions\label{conclusions-sect}}

In this paper, we explore the power of combined multi-wavelength observations to unveil details of protoplanetary disks that remain hidden, or 
are degenerated, when observed with a single technique. We show that combinations of observations can be used to remove the degeneracy
in the interpretation of the disk structure, to distinguish disks with different structures, and to explore the paths of disk evolution.
Well-devised, combined observations, even those with relatively old facilities, can help to unveil details of disks that are beyond the reach
even of the most modern telescopes, also contributing to make the most with current ground- and space-based instrumentation. 

The power of multiwavelength
data is also an indication that future advances will need to involve many experts working on different techniques.
Our proposed Disk's Rosetta Stone explores useful combinations of
observations depending on the particular problem that we want to address, helping to devise time- and telescope-efficient observing schemes.  
The Disk's Rosetta Stone is an ever-growing scheme: our study is not
complete and disk decryption will improve as new observations become feasible and gain sensitivity, power, efficiency, and time baseline.


\begin{acknowledgements}
We thank the organizers of the Protoplanetary Discussions in Edinburgh, March 2016, and in particular, 
P. Woitke, for bringing all of us together and making possible this observational collaboration, and also for his comments to the manuscript. 
We also thank M. Benisty for her comments and suggestions and her participation in the discussion, D. Price for his support of this paper, B. L\'{o}pez
for his input regarding MIDI, and the
anonymous referee for valuable comments that contributed to clarify this work.
This work is the fruit of a collaborative effort between disk observers that use different observing techniques. A.S.A. is the organizer of the paper
contributing to Sections \ref{innerdisk-sect}, \ref{accretion-sect}, \ref{timeresolved-sect}. She also organized Section \ref{discussion-sect}, to which all
authors have contributed. A.B. is the co-organizer of the paper, leading Sections \ref{innerdisk-sect} and \ref{gaps-sect} and contributing to Section \ref{discussion-sect}
and to the overall structure of the paper.
A.C. and M.K. contributed to Sections \ref{innerdisk-sect} and \ref{gaps-sect}. M.K is also the main contributor to Section \ref{diskmass-sect}. 
T. S. and A.G. led the scattering light and polarization observations in Section \ref{gaps-sect}. I.M. is the main contributor to accretion in 
massive stars in Section \ref{accretion-sect} and also
contributed to Section \ref{timeresolved-sect}. K.F. contributed to the disk dynamics part in Section \ref{timeresolved-sect}. N.v.d.M. is the main contributor to the mm continuum
interferometry part in Section \ref{gaps-sect}. J.G. contributed to the dust mass part in
Section \ref{diskmass-sect}. All authors contributed to the discussion in Section \ref{discussion-sect}, as well as in building the Disk's Rosetta Stone in Figures \ref{rosetta-fig}
and \ref{feasibility-fig}.
This research has made use of NASA's Astrophysics Data System.

\end{acknowledgements}

\def\nar{New A Rev.}%

\bibliographystyle{pasa-mnras}
\bibliography{1r_lamboo_notes}

\nocite*{}

\end{document}